\documentclass[aps,pre,reprint]{revtex4-1}
\pdfoutput=1
\usepackage[utf8x]{inputenc}
\usepackage{amsmath,amsfonts,amssymb}
\usepackage{graphicx}
\usepackage{amsthm}

\usepackage{algpseudocode}
\usepackage{algorithm}
\usepackage{tikz-cd}

\newtheorem{theorem}{Theorem}
\newtheorem{lemma}{Lemma}

\begin{document}
\title{Solutions of the Multivariate Inverse Frobenius--Perron Problem}

\date{\today}
\author{Colin Fox}
\affiliation{Department of Physics, University of Otago, Dunedin, New Zealand. email:colin.fox@otago.ac.nz}
\author{Li-Jen Hsiao}
\affiliation{System Manufacturing Center, National Chung-Shan Institute of Science \& Technology, Taiwan. email:ljh.lijenhsiao@gmail.com}
\author{Jeong Eun (Kate) Lee}
\affiliation{Department of Statistics, The University of Auckland, Auckland, New Zealand. email:kate.lee@auckland.ac.nz}


\begin{abstract}
We address the inverse Frobenius--Perron problem: given a prescribed target distribution $\rho$, find a deterministic map $M$ such that iterations of $M$ tend to $\rho$ in distribution. We show that all solutions may be written in terms of a factorization that combines the forward and inverse Rosenblatt transformations with a uniform map, that is, a map under which the uniform distribution on the $d$-dimensional hypercube as invariant. Indeed, every solution is equivalent to the choice of a uniform map. We motivate this factorization via $1$-dimensional examples, and then use the factorization to present solutions in $1$ and $2$ dimensions induced by a range of uniform maps.
\end{abstract}

\maketitle

\section{Introduction}
A basic question in the theory of discrete dynamical systems, and in statistical mechanics, is whether a chaotic iterated function $M:X\rightarrow X$, that maps a space $X\subseteq\mathbb{R}^d$ back onto $X$, has an equilibrium distribution, with probability density function (PDF) $\rho(x)$~\footnote{The density function is with respect to some underlying measure, typically Lebesgue. Throughout this paper we will use the same symbol for the distribution and associated PDF, with meaning taken from context.}. A necessary condition is that the distribution $\rho$ is \emph{invariant} under $M$, i.e. if $x\sim \rho$ ($x$ is distributed as $\rho$) then so is $M(x)$, and further that the orbit of almost all points $x\in X$ defined as $\mathcal{O}^+(x)=\left\{x,M(x),M^2(x),M^3(x),\ldots\right\}$ tends in distribution to $\rho$. 
Then, under mild conditions, the map is ergodic for $\rho$, that is, expectations with respect to $\rho$ may be replaced by averages over the orbit~\cite{dorfman1999introduction,LasotaMackey}. 

For example, it is well known~\cite{UlamvonNeumann1947,may1976simple,dorfman1999introduction,LasotaMackey}  that the logistic map $M_\text{log}(x)=4x(1-x)$, for $x\in [0,1]$, is chaotic with the equilibrium distribution having PDF
\[ \rho_\text{log}(x) = \frac{1}{\pi \sqrt{x(1-x)}}, \]
implying that $M_\text{log}$ is ergodic for $\rho_\text{log}$.

Our motivating interest is in using this ergodic property to implement sample-based inference for Bayesian analysis or machine learning. In those settings, the target distribution $\rho$ is defined by the application. Generating a sequence $\left\{x_0,x_1,x_2,x_3,\ldots\right\}$ that is ergodic for $\rho$ is useful because then  expectations of any quantity of interest can be computed as averages over the sequence, i.e.,
\begin{equation}
 \lim_{N\to\infty}\frac{1}{N}\sum_{i=0}^{N-1} g\left(x_i\right) =\int_X g(x)\rho(x)\,\mathrm{d} x .
 \label{eq:expect}
\end{equation}
Commonly in statistics, such ergodic sequences are generated using a \emph{stochastic} iteration that simulates a Markov chain targeting $\rho$~\cite{mcmc_liu}; here we explore \emph{deterministic} iterations that generate an orbit that is ergodic for $\rho$.

The equilibrium distribution for a given iterated function, if it exists, can be approximated by computing the orbit of the map for some starting point then performing kernel density estimation, or theoretically by seeking the stationary distribution of the  Frobenius--Perron (FP) operator that is the transfer~\cite{KlusKoltaiSchutte2016}, aka push-forward, operator induced by a deterministic map~\cite{dorfman1999introduction,LasotaMackey}; we present the FP  equation in Section~\ref{sec:LEFP}. 

The inverse problem that we consider here, of determining a map that gives a prescribed equilibrium distribution, is the \emph{inverse Frobenius--Perron problem} (IFPP) and has been studied extensively~\cite{grossmann1977invariant,ershov1988solution,diakonos1996construction,diakonos1999stochastic,pingel1999theory,bollt2000controlling,nie2013new,nie2018matrix}. Summaries of previous approaches to the IFPP are presented in~\cite{bollt2000controlling,rogers2008synthesis} that characterize approaches as based on conjugate functions (see \cite{grossmann1977invariant} for details) or the matrix method (see~\cite{rogers2008synthesis} for details), and~\cite{wei2015solutions} that also lists the differential equation approach. Existing work almost solely considers the IFPP in one-variable, $d=1$, the exception being the development of a two-dimensional solution in~\cite{bollt2000controlling} that is also presented in~\cite{rogers2008synthesis}.

The matrix method, first suggested by Ulam~\cite{Ulam1960}, solves the IFPP for a piecewise-constant approximation to the target density, using a transition matrix representation of the approximated FP operator. Convergence of the discrete approximation is related to Ulam's conjecture, and has been proved for the multidimensional  problem; see~\cite{wei2015solutions} and references therein. While the matrix method allows construction of solutions, at least in the limit, existing methods only offer a limited class of very non-smooth solutions, so are not clearly useful for characterizing all solutions, as we do here. We do not further consider the matrix methods.

The development in this paper starts with the differential equation approach in which the IFPP for restricted forms of distributions and maps is written as a differential equation that may be solved. We re-derive some existing solutions to the IFPP this way in Sec.~\ref{sec:IFPP}. The main contribution of this paper is to show that the form of these solutions may be generalized to give the general solution of the IFPP for any probability distribution in any dimension $d$, as presented in the factorization theorem of Sec.~\ref{sec:fact}.  This novel factorization represents solutions of the IFPP in terms of the Rosenblatt transformation~\cite{rosenblatt-1952} and a \emph{uniform map}, that is a map on $[0,1]^d$ that leaves the uniform distribution invariant. In particular, we show that the conjugating functions in~\cite{grossmann1977invariant} are exactly the inverse Rosenblatt transformations. For a given Rosenblatt transformation, there is a one-to-one correspondence between solution of the IFPP and choice of a uniform map. 

This reformulation of the IFPP in terms of two well-studied constructs leads to practical analytic and numerical solutions by exploiting existing, well-developed methods for Rosenblatt transformations, and for deterministic iterations that target the uniform distribution. The factorization also allows us to establish equivalence of solutions of the IFPP and other methods that employ a deterministic map within the generation of ergodic sequences.  This standardizes and simplifies existing solution methods by showing that they are special cases of constructing the Rosenblatt transformation (or its inverse), and selection of a uniform map.

This paper starts with definitions of the IFPP and the Lyuapunov exponent in Sec.~\ref{sec:LEFP}. Solutions of the IFPP in one-dimension, $d=1$, are developed in Sec.~\ref{sec:IFPP}. These solutions for $d=1$ motivate the factorization theorem in Sec.~\ref{sec:fact} that presents a general solution to the IFPP for probability distributions with domains in $\mathbb{R}^d$ for any $d$. Sec.~\ref{sec:onedim}  presents further examples of univariate, $d=1$, solutions to the IFPP based on the factorization theorem  in Sec.~\ref{sec:fact}.
Two 2-dimensional numerical examples are presented in Section~\ref{sec:num} to demonstrate that the theoretical constructs may be implemented in practice. A summary and discussion of results is presented in Section~\ref{sec:sad}, including a discussion of some existing computational methods that can be viewed as implicitly implementing the factorization solution of the IFPP presented here.

\section{Inverse Frobenius--Perron Problem and Lyapunov Exponent}
\label{sec:LEFP}

In this section we define the forward and inverse Frobenius--Perron problems, and also the Lyuapunov exponent that measures chaotic behaviour. 

\subsection{Frobenius--Perron Operator}
\label{sec:FPop}
A deterministic map $x_{n+1} = M(x_n)$ defines a map on probability distributions over state, called the transfer operator~\cite{KlusKoltaiSchutte2016}. Consider the case where the initial state $x_0\sim\rho_0(\cdot)$ ($x_0$ is distributed as $\rho_0$) for some distribution $\rho_0$ and let $\rho_n$ denote the $n$-step distribution, i.e., the distribution over $x_n=M^n(x_0)$ at iteration $n$. The transfer operator that maps $\rho_{n}\mapsto\rho_{n+1}$ induced by $M$  is given by the Frobenius--Perron operator associated with $M: x\mapsto y$~\cite{dorfman1999introduction,LasotaMackey,Gaspard1998}
\begin{equation}
  \rho_{n+1}(y) = \sum_{x\in M^{-1}(y)} \frac{\rho_{n}\left(x \right)}{|J(x)|}
  \label{eq:FP}
\end{equation}
where $|J(x)|$ denotes the Jacobian determinant~\footnote{We have used the language of differential maps, as all the maps that we display in this paper are differentiable almost everywhere~\cite{varberg1971change}. More generally, $|J(x)|^{-1}$ denotes the density of $\rho_{n} M^{-1}$ with respect to $\rho_{n+1}$, see, e.g., \cite[Remark 3.2.4.]{LasotaMackey}.} of $M$ at $x$, and the sum is over inverse images of $y$.

The equilibrium distribution $\rho$ of $M$ satisfies 
\begin{equation} \label{eq:FPinvar} 
  \rho(y) = \sum_{x\in M^{-1}(y)} \frac{\rho(x)}{|J(x)|} 
\end{equation}
and we say that $\rho$ is invariant under $M$.

\subsection{Inverse Frobenius--Perron Problem}

The inverse problem that we address is finding an iterative map $M$ that has a given distribution $\rho$ as its equilibrium distribution. We do this by performing the inverse Frobenius--Perron problem (IFPP) of finding $M$ that satisfies~\eqref{eq:FPinvar} to ensure that $\rho$ is an invariant distribution of $M$. Establishing chaotic hence ergodic behaviour is a separate calculation.

We will assume throughout that $\rho$ is absolutely continuous with respect to the underlying measure so that a probability density function $\rho(x)$ exists, and, furthermore, that $\rho(x)>0$, $\forall x\in X$.

\subsection{Lyapunov Exponent}

The Lyapunov exponent $h$ of an iterative map gives the average exponential rate of divergence of trajectories.
We define the (maximal) Lyapunov exponent $h$ as~\cite{dorfman1999introduction,LasotaMackey}
\begin{equation}
   h = \lim_{N\to\infty} \frac{1}{N} \log \left|\frac{\mathrm{d} x_N}{\mathrm{d} x_0}\right|
   = \lim_{N\to\infty}\frac{1}{N} \sum^{N-1}_{n=0} \log|J(x_n)| 
   \label{eq:Lyap}
\end{equation}
that features the starting value $x_0$. For ergodic maps the dependency on $x_0$ is lost as $N\to\infty$  and the Lyapunov exponent may be written  
\begin{equation}
  h = \int_X \log|J(x)| \rho(x)\, \mathrm{d} x \label{eq:LyapExpect}
\end{equation}
where $\rho(x)$ is the invariant density. A positive Lyapunov  exponent $h$ indicates that the map is chaotic.

The theoretical value for the Lyapunov exponent may be obtained using~\eqref{eq:LyapExpect}, while~\eqref{eq:Lyap} provides an empirical value obtained by iterating the map $M$. For example, the Lyapunov exponent of the logistic map evaluated by~\eqref{eq:LyapExpect} is $h_\text{log} = \log 2\approx 0.693147$ while~\eqref{eq:Lyap} evaluated over an orbit with $10000$ iterations gave $h_\text{log}\approx 0.693140$.

For chaotic maps, any uncertainty in initial value  means that an orbit cannot be precisely predicted, since initial states with any separation become arbitrarily far apart, within $X$, as iterations increase. Hence it is useful to characterize the orbit statistically, in terms of the equilibrium distribution over states in the orbit.

It is interesting to note that theoretical chaotic and ergodic behaviour does not necessarily occur when iterations of a map are implemented on a finite-precision computer. For example, when the logistic map, above, on $[0,1]$ is iterated on a binary computer the multiplication by $4$ corresponds to a shift left by two bits and all subsequent operations maintain lowest order bits that are zero. Repeated iterations eventually produce the number zero, no matter the starting value. While it is simple to correct this non-ideal behaviour, as was done to give the numerical Lyapunov exponent, above, it is important to note that computer implementation can have very different dynamics to the mathematical model.

\section{Solution of the IFPP in $1$-dimension}
\label{sec:IFPP}

In this Section we develop solutions of the IFPP in one-dimension, $d=1$. Without loss of generality we consider distributions on the unit interval $X=[0,1]$, as the domain of any univariate distribution may be transformed by a change of variable to $X=[0,1]$, including when the domain is the whole real line $(-\infty,\infty)$.

For distributions over a scalar random variable, the FP equation for the invariant density~\eqref{eq:FPinvar} simplifies to
\begin{equation}
  \rho(y) = \sum_{x\in M^{-1}(y)} \frac{\rho\left(x \right)}{|M'(x)|}.
  \label{eq:FP1dim}
\end{equation}

\subsection{The Simplest Solution}
\label{sec:simplest}
We first note, almost trivially, that the identity map $M=I$, where $I(x)=x$, has $\rho$ as an invariant distribution, so solves the IFPP for any $\rho$. Somewhat less trivial is to derive this simplest solution by assuming that $M$ is monotonic increasing and $M(0)=0$, so that there is just one inverse image in~\eqref{eq:FP1dim}. Writing $|M'(x)|=\mathrm{d} M/\mathrm{d} x$ gives the differential equation with separated variables
\begin{equation}
  \rho(M) \mathrm{d} M = \rho(x) \mathrm{d} x 
  \label{eq:sepde}
\end{equation}
that has solution
\[ F(M) = F(x) \]
where $F(x)=\int_0^x\rho(x')\,\mathrm{d} x'$ is the cumulative distribution function (CDF) for $\rho$.
If $F$ is invertible denote the inverse by $F^{-1}$, called the the inverse distribution function (IDF), otherwise let $F^{-1}$ denote the generalized inverse distribution function, 
$ \displaystyle F^{-1}(p)=\inf\{x\in X :F(x)\geq p\}$. Then, $M=F^{-1}(F(x)) = x$, or $M(x)=x$, almost everywhere. Hence the identity map is the \emph{unique} monotonic increasing map that has $\rho$ as its invariant distribution. Clearly, the identity map is not ergodic for $\rho$. 

\begin{figure} \centering
     \includegraphics[width=0.49\linewidth]{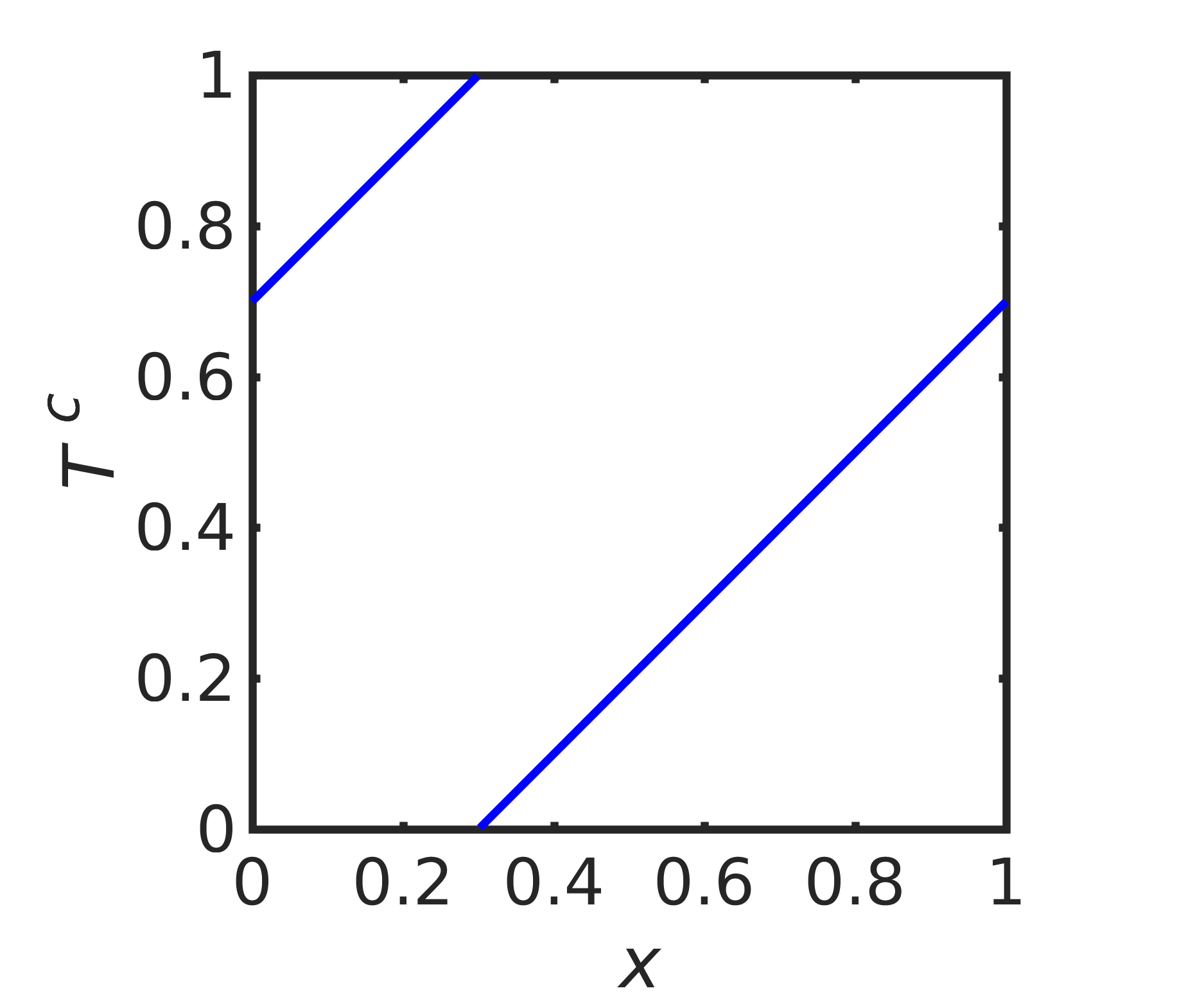}
     \includegraphics[width=0.49\linewidth]{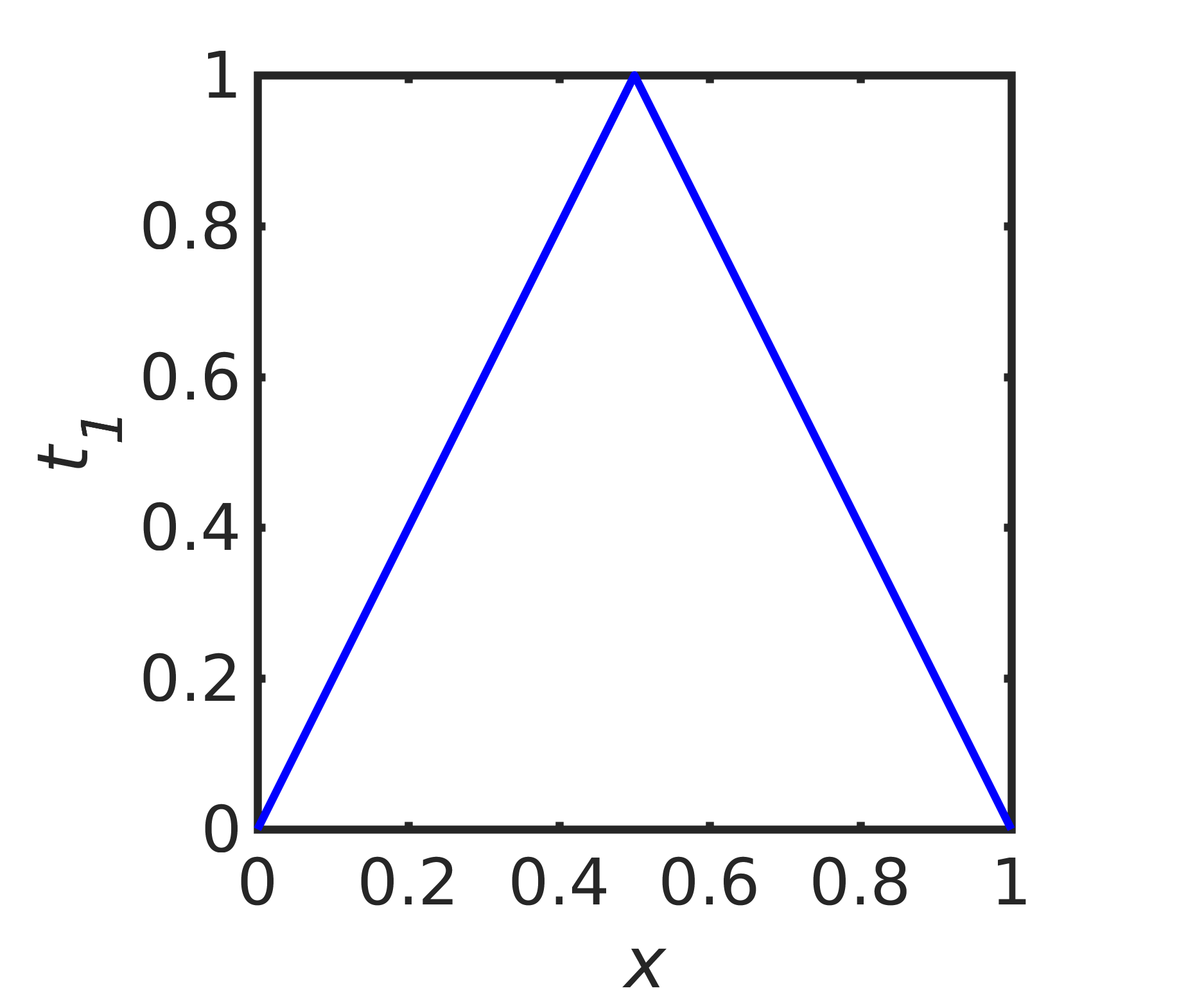}
     \caption{Translation operator $T^c$ for $c=0.3$ (left) and triangle map $t_1$ (right).}
     \label{fig:Tct1}
\end{figure}

We may generalize this solution by setting $M(0)= k$, for some $k\in [0,1)$, and also only requiring $M$ to be piecewise continuous. Allowing one discontinuity in $M$ we write the integral of the separated differential equation~\eqref{eq:sepde} as
\[ F(M) = F(x) +k  \mod 1 \]
giving the solution to the IFPP
\begin{equation}
  M(x) = \left(F^{-1} \circ T^c \circ F\right)(x)
  \label{eq:shift}
\end{equation}
where $c=F^{-1}(k)$. Here $T^c$ denotes the operator that translates by $c$ with wrap-around on $[0,1)$~\footnote{Hence $T^c$ is the translation operator on the unit circle $S^1$} 
\begin{equation}
 T^c(y) = y + c - \lfloor y + c \rfloor ,
 \label{eq:Tc}
\end{equation}
where $\displaystyle \lfloor \,\ \rfloor $ denotes the floor function; see Figure~\ref{fig:Tct1} (left). The identity map is recovered when $k=c=0$.

It is easily seen that the Lyapunov exponent of the map in~\eqref{eq:shift} is $h=0$, so the map is not chaotic. However, interestingly, this map can generate a sequence of states that produce a numerical integration rule with respect to $\rho$, since appropriate choices of $c$ and number of iterations $N$ can produce a rectangle rule quadrature or a quasi Monte Carlo lattice rule; see Sec.~\ref{sec:prop}.

\subsection{Exploiting Symmetry in $\rho(x)$}
When the PDF $\rho$ has reflexive symmetry about $1/2$, i.e. $\rho(x)=\rho(1-x)$, we can simplify the FP equation~\eqref{eq:FP1dim} by assuming that the map $M$ has the same symmetry. Specifically, we write the triangle map (see Figure~\ref{fig:Tct1} (right))
\begin{equation}
 t_1(x) = 1-2\left|x-1/2\right|
 \label{eq:t1}
\end{equation}
that has reflexive symmetry about $1/2$, and write 
\begin{equation}
 M(x) = m(t_1(x))
 \label{eq:doubsym}
\end{equation}
where $m(x):[0,1]\to [0,1]$ is a monotonic increasing map with $m(0)=0$ (and, it will turn out, $m(1)=1$). 
Hence the FP equation simplifies to
\begin{equation}
  \rho(y) = 2\frac{\rho\left(x \right)}{|M'(x)|}, \quad  x \in M^{-1}(y),
  \label{eq:FPsym}
\end{equation}
that we can write as the separated equations
\begin{eqnarray*}
  \rho(M)\, \mathrm{d} M = 2 \rho(x) \mathrm{d} x, \quad & x<1/2, \\
  \rho(M)\, \mathrm{d} M = -2 \rho(x) \mathrm{d} x, \quad & x>1/2,
\end{eqnarray*}
that has the continuous solution
\[ F(M) = t_1(F(x)) \]
giving the continuous solution to the IFPP
\begin{equation}
  M(x) = \left(F^{-1} \circ t_1 \circ F\right)(x).
  \label{eq:sym}
\end{equation}
One can solve for $m(x) = F^{-1}(2F(x/2))$, though we do not further consider the function $m$.

The approach we have used here simplifies the approach in~\cite{diakonos1996construction}, while  `doubly symmetric' maps of the form~\eqref{eq:doubsym} were considered in~\cite{gyorgyi1984fully}, and again in~\cite{pingel1999theory,diakonos1999stochastic}.

\subsection{Symmetric Triangular Distribution}
\label{sec:tent}
\begin{figure} \centering
     \includegraphics[width=0.49\linewidth]{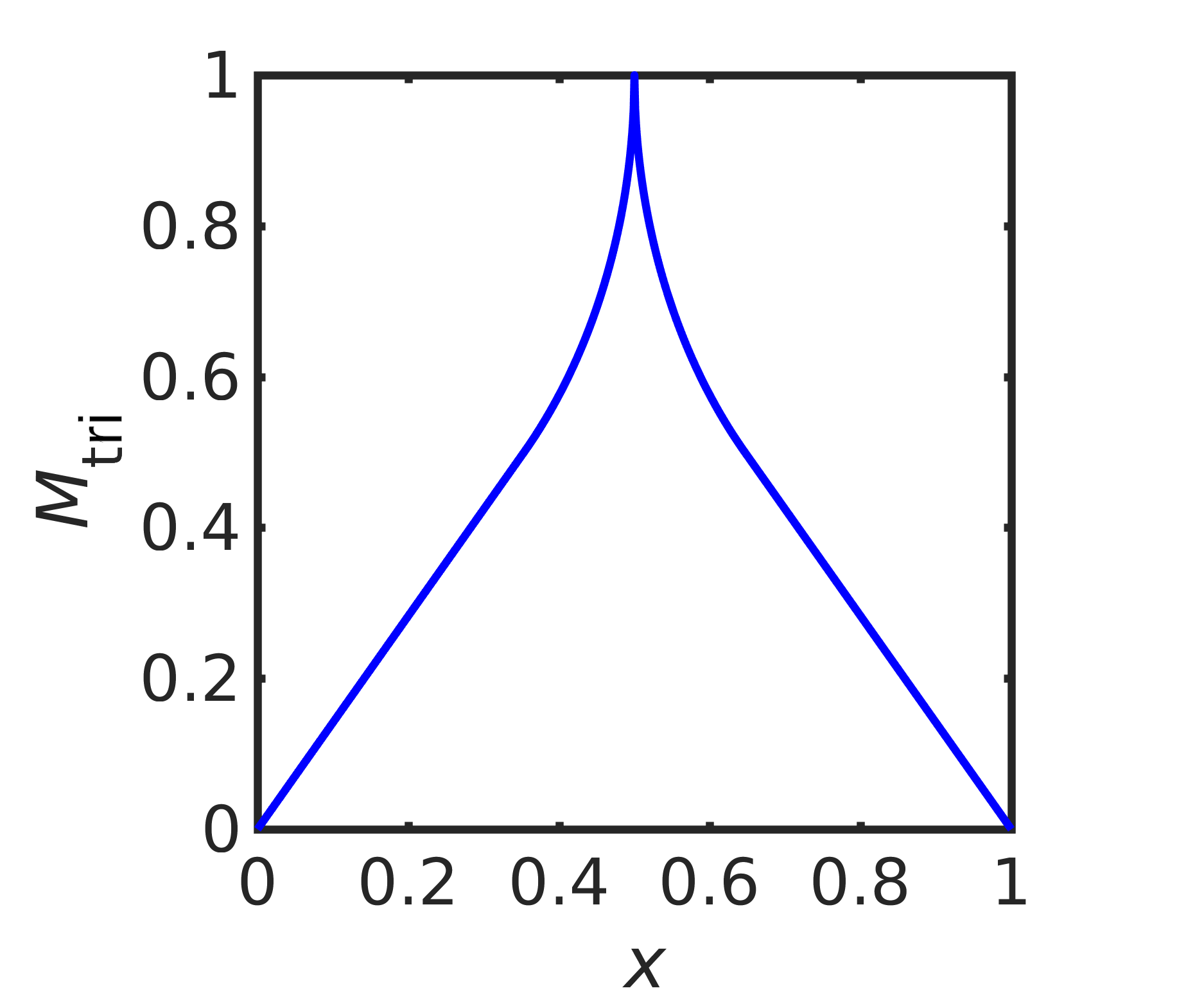}
     \includegraphics[width=0.49\linewidth]{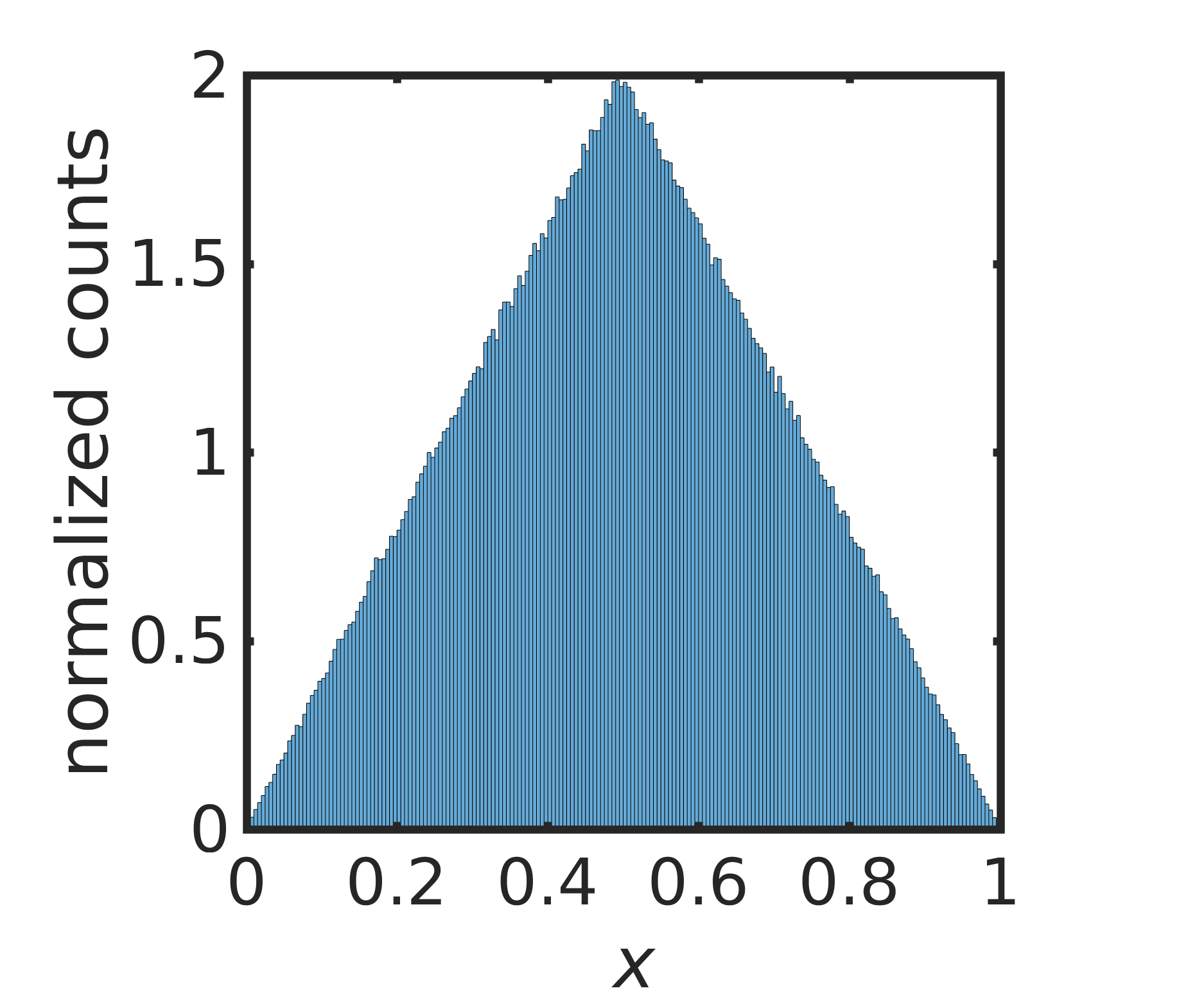}
     \caption{Iterative map $M_\text{tri}$ in~\eqref{eq:Mtent} (left) and a histogram of $1\times 10^6$ iterates of the map $M_\text{tri}$ (right).}
     \label{fig:Mtent}
\end{figure}

To give a concrete example of the solution in~\eqref{eq:sym}, we consider the symmetric triangular distribution on $[0,1]$ with PDF
\begin{equation}
 \rho_\text{tri}(x) = 2-4\left| x-\frac{1}{2} \right|
\end{equation}
that has reflexive symmetry about $x=\frac{1}{2}$. The CDF is 
\[ F_\text{tri}(x) = \begin{cases}
                        2x^2                    & 0\le x \le \frac{1}{2} \\
                        \frac{1}{2} + 4x - 2x^2 & \frac{1}{2} \le x \le 1
                      \end{cases}
\]
giving the unimodal map, after substituting into~\eqref{eq:sym}, 
\begin{equation}
  \label{eq:Mtent}
  M_\text{tri}(x) = \left\{ \begin{array}{rl}
  \sqrt{2}x & 0\le x \le \frac{1}{\sqrt{8}} \\
  1-\sqrt{\frac{1}{2}-2x^2} & \frac{1}{\sqrt{8}} \le x \le \frac{1}{2} \\
  1-\sqrt{\frac{1}{2}-2(1-x)^2} & \frac{1}{2} \le x \le 1 - \frac{1}{\sqrt{8}} \\
  \sqrt{2}(1-x) & 1 - \frac{1}{\sqrt{8}} \le x \le 1
  \end{array} \right.
 \end{equation} 
shown in Figure~\ref{fig:Mtent} (left). The same map was derived in~\cite{grossmann1977invariant}. Figure~\ref{fig:Mtent} (right) shows a normalized histogram of $10^6$ iterates of $M_\text{tri}$ starting at $x=0.3$, confirming that the orbit of $M_\text{tri}$ converges to the desired triangular distribution. The numerical implementation avoids finite-precision effects, as discussed later.

The theoretical Lyapunov exponent for $M_\text{tri}$ is $h_\text{tri}=\log 2\approx 0.693147$ while~\eqref{eq:Lyap} evaluated over an orbit with $10^6$ iterations gave $h_\text{tri}\approx 0.693148$.

\section{Solutions of the IFPP for General Multi-Variate Target Distributions}
\label{sec:fact}
The solutions to the $1$-dimensional IFPP with special structure in Eqs~\eqref{eq:shift} and \eqref{eq:sym} are actually examples of a general solution to the IFPP for multi-variate probability distributions with no special structure. We state that connection via a theorem that establishes a factorization of all possible solutions to the IFPP, and that also provides a practical means of solving the IFPP. 

We first introduce the forward and inverse Rosenblatt transformations, that is the multi-variate generalization the CDF and IDF for univariate distributions.

\subsection{Forward and Inverse Rosenblatt Transformations}
\label{sec:rosen}
A simple transformation of an absolutely continuous $d$-variate distribution into the uniform distribution on the $d$-dimensional hypercube was introduced by Rosenblatt~\cite{rosenblatt-1952}, as follows. 
The joint PDF can be written as a product of conditional densities,
\[
 \rho(x_1,\ldots,x_d) = \rho_1(x_1) \rho_2(x_2|x_1) \cdots \rho_d(x_d | x_1\ldots,x_{d-1}),
\]
where $\rho_k(x_k| x_1\ldots,x_{k-1})$ is a conditional density given by
\begin{equation}
\rho_k(x_k| x_1\ldots,x_{k-1}) = \frac{p_k(x_1,\ldots,x_k)}{p_{k-1}(x_1\ldots,x_{k-1})},
\label{eq:rho_k}
\end{equation}
in terms of the marginal densities,
\begin{equation}
 p_k = \int \rho(x_1, \ldots,x_d)\, \mathrm{d} x_{k+1} \cdots \mathrm{d} x_d,
 \label{eq:mdf}
\end{equation}
where $k=1,\ldots,d$.

Let $z=(z_1,\ldots,z_d)=R(x_1,\ldots,x_d)$ where $R$ is the Rosenblatt transformation~\cite{rosenblatt-1952} from the state-space $X \subseteq\mathbb{R}^d$ of $\rho$ to the $d$-dimensional unit cube $[0,1]^d$, defined in terms of the (cumulative) distribution function $F$ by
\begin{align*}
 z_1 &=  F_1(x_1) = \int_{-\infty}^{x_1}\rho_1(x_1')\, \mathrm{d} x_1' ,\\
 z_2 &=  F_2(x_2|x_1) = \int_{-\infty}^{x_2}\rho_2(x_2'|x_1)\, \mathrm{d} x_2' ,\\
     &\vdots \\
 z_d &=  F_d(x_2|x_1,\ldots,x_{d-1}) = \int_{-\infty}^{x_d}\rho_d(x_d'|x_1,\ldots,x_{d-1})\, \mathrm{d} x_d' .
\end{align*}

As noted in~\cite{rosenblatt-1952}, there are $d!$ transformations of this type, corresponding to the $d!$ ways of ordering the coordinates. Further multiplicity is introduced by considering coordinate transformations, such as rotations. 

Notice that in $1$-dimension, the Rosenblatt transformation $R(x)$ is simply the CDF $F(x)$.
                                                                                                                                                                                                        
It follows that if $x\sim \rho$ then $z=R(x)\sim \operatorname{Unif}([0,1]^d)$, i.e., $z$ is uniformly distributed on the $d$-dimensional unit cube~\cite{rosenblatt-1952}. When $\rho(x)>0$, $\forall x\in X$ the distribution functions are strictly monotonic increasing and the inverse of the Rosenblatt transformation $R^{-1}$ is well defined, otherwise let $R^{-1}$ denote the generalized inverse as in Sec.~\ref{sec:simplest}. Then, if $z\sim \operatorname{Unif}([0,1]^d)$ it follows that $x= R^{-1}(z)\sim \rho$, i.e., $x$ is distributed as the desired target distribution $\rho$~\cite{johnson1987multivariate}; this is the basis of the \emph{conditional distribution method} for generating multi-variate random variables, that generalizes the inverse cumulative transformation method for univariate distributions~\cite{devroye-rvgen-1986,johnson1987multivariate,hormann-rvgen-2004,dolgov2020approximation}. These results may also be established by substituting $R$ or $R^{-1}$ into the the FP equation~\eqref{eq:FP}, noting that there is a single inverse image and that the Jacobian determinant of $R$ equals the target PDF $\rho(x)$.  

In the remainder of this paper, we refer to \emph{any} map $R$ satisfying $x\sim \rho \Rightarrow R(x)\sim \operatorname{Unif}([0,1]^d)$ as a Rosenblatt transformation, with (generalized) inverse as defined above. 

\subsection{Factorization Theorem}
The following theorem characterizes solutions to the IFPP.
\begin{theorem}
\label{theo:fact}
 Given a probability distribution $\rho$ in $d$-dimensions, a map  $M(x)$ is a solution of the IFPP, that is, $M(x)$ satisfies the FP equation~\eqref{eq:FPinvar}, if and only if
 \begin{equation} 
   M(x) = (R^{-1} \circ U \circ R)(x),
   \label{eq:fact} 
 \end{equation}
 where $R$ is a Rosenblatt transformation and $U$ is a `uniform map' on the unit $d$-dimensional hypercube, i.e. a map that has $\operatorname{Unif}([0,1]^d)$ as invariant distribution~\footnote{When $U$ satisfies the stronger condition that $\operatorname{Unif}([0,1]^d)$ is the equilibrium distribution, $U$ is called an exact map~\cite{LasotaMackey}.}. 
\end{theorem}
\noindent \textbf{Proof:}
We will show that $\rho$ is an invariant distribution of $M$ iff $M$ has the form~\eqref{eq:fact}. ($\Rightarrow$) Assume $M$ has the form~\eqref{eq:fact}. If $x\sim\rho$ then $R(x)\sim \operatorname{Unif}([0,1]^d)$ hence $U(R(x))\sim \operatorname{Unif}([0,1]^d)$, as $\operatorname{Unif}([0,1]^d)$ in invariant under $U$, hence  $M(x)=R^{-1}(U(R(x)))\sim\rho$, as desired. ($\Leftarrow$) If $\rho$ is invariant under $M$ then 
\( U = R \circ M \circ R^{-1} \)
is a uniform map, since if $z\sim \operatorname{Unif}([0,1]^d)$ then $R^{-1}(z)\sim\rho$, $M(R^{-1}(z))\sim\rho$, and $R(M(R^{-1}(z)))\sim\operatorname{Unif}([0,1]^d)$. Inserting this $U$ into~\eqref{eq:fact} gives the desired factorization. 
\quad\qed

The first part of the proof shows that any uniform map $U$ induces a solution to the IFPP, though the particular solution depends on the particular Rosenblatt transformation. The second part of the proof shows that different solutions to the IFPP effectively differ only by the choice of the uniform map $U$, once the Rosenblatt transformation is determined, that is, a coordinate system is chosen with an ordering of those coordinates.

Grossmann and Thomae~\cite{grossmann1977invariant} called dynamical systems $M$ and $U$ related by a formula of the form~\eqref{eq:fact} as `related by conjugation', or just `conjugate', and the map $R^{-1}$ in~\eqref{eq:fact} is a `conjugating function'. Thus, in the language of~\cite{grossmann1977invariant}, Theorem~\ref{theo:fact} shows that the IFPP for any distribution $\rho$ has a solution (actually, it shows that there are infinitely many solutions), every solution map is conjugate to a uniform map, and the conjugating function is precisely the inverse Rosenblatt transformation. 

Notice that both the translation operator $T^c$ in~\eqref{eq:Tc} and the triangle map in~\eqref{eq:t1} are uniform maps on the unit interval $[0,1]$. Thus, the solutions to the IFPP given in Eqs~\eqref{eq:shift} and \eqref{eq:sym} are examples of the general solution form in~\eqref{eq:fact}. In particular, while the solution to the IFPP in~\eqref{eq:sym} was derived assuming symmetry of the target density $\rho(\cdot)$, \eqref{eq:sym} actually gives a solution of the IFPP for \emph{any} density $\rho(\cdot)$. Unimodal maps of this form were derived in~\cite{ershov1988solution}.

Computed examples of solutions to the IFPP given by the factorization~\eqref{eq:fact} are presented in Sec.~\ref{sec:onedim}, in one dimension, and in Sec~\ref{sec:twodim} in two dimensions. High dimensional calculations are discussed in Sec.~\ref{sec:sad}.

\subsection{Properties of $M$ from $U$}
\label{sec:prop}
Many properties of the map $M$ are inherited from the uniform map $U$. 

When $R$ and $R^{-1}$ are continuous, $M$ is continuous iff $U$ is continuous. In one dimension, monotonicity of the CDF and IDF imply that the number of modes of $U$ equals the number of modes of $M$; in particular, $M$ is unimodal iff $U$ is unimodal. 

Constructing iterative maps with specific periodicity of the orbit is possible through the use of translation operators $T^c$ as uniform maps, defined in Eq.~\eqref{eq:Tc}. Consider, first, maps in one dimension on $[0,1]$.
If the shift $c\neq0$ and $c\notin\mathbb{Q}$, the map is aperiodic. However, in the case that $c\neq0$ and $c\in\mathbb{Q}$ such that
 \begin{equation} c = \frac{N}{D} \end{equation}
with $N,D\in\mathbb{N}$ and $\gcd(N,D)=1$, then the map is periodic with periodicity $D$, and iterative maps constructed with $U=T^c$ exhibit the same periodicity. These properties may be extended to multi-dimensional settings when the translation constant $c$ is a vector of shifts in each coordinate direction, as used in rank-one lattice rules for quasi-Monte Carlo integration~\cite{dick2013high}. 

The factorization in Theorem~\ref{theo:fact} also shows that performing an iteration $x_{n+1}=M(x_n)$ with an iterative map $M$ on the space $X$  is equivalent to applying the corresponding uniform map $z_{n+1}=U(z_n)$ on the space $[0,1]^d$ through the transformations $R$ and $R^{-1}$, as indicated in the following (commuting) diagram.
\[
 \begin{tikzcd}
   x_n \arrow[r, "\displaystyle R" ']  \arrow[d, left, "\displaystyle M" ']  & z_n \arrow[d, "\displaystyle U" ] \arrow[l, shift right=1.5ex, "\displaystyle R^{-1}" '] \\
   x_{n+1} \arrow[r,  shift right=1.5ex, "\displaystyle R" ']  & z_{n+1} \arrow[l, "\displaystyle R^{-1}" '] 
 \end{tikzcd}
\]
Using this commuting property, it is straightforward to prove the following lemma.
\begin{lemma} \label{lem:commute}
 For given distribution $\rho$, let $R$ be a Rosenblatt transformation for $\rho$. 
 Let $M=R^{-1}\circ U\circ R$ be a solution of the IFFP, as guaranteed by Theorem~\ref{theo:fact}, where $U$ is a uniform map. 
 Then
 \begin{equation}
  \mathcal{O}^+_M(x_0)= R^{-1}\left(\mathcal{O}^+_U(R(x_0)) \right).
  \label{eq:commute}
 \end{equation}
\end{lemma}
Hence, instead of iterating $M$ on the space $X$ to produce the sequence $\{x_1,x_2,x_3,\ldots\}$, Eq.~\eqref{eq:commute} shows that one can iterate the map $U$ on the space  $[0,1]^d$ to produce the sequence $\{z_1,z_2,z_3,\ldots\}$ and then evaluate $x_n=R^{-1}z_n$, $n=1,2,\ldots$, to produce exactly the same sequence on $X$. Since the map $M$ is mixing or ergodic iff the uniform map $U$ is mixing or ergodic, respectively, in this sense mixing and ergodicity of $M$ is inherited from $U$. 

Using the expansion in Eq.~\eqref{eq:commute}, we see that $M$ is deterministic or stochastic iff $U$ is deterministic or stochastic, respectively. Even though we are not considering stochastic maps here, we note that, for stochastic maps, iterations of $M$ are correlated or independent iff iterations of $U$ are correlated or independent, respectively.

Some other properties that are, and are not, preserved by the transformation from $U$ to $M$ are discussed in~\cite{grossmann1977invariant}.

\section{Examples in One Dimension}
\label{sec:onedim}

\subsection{Uniform Maps on $[0,1]$}
\begin{figure} \centering
     \includegraphics[width=0.49\linewidth]{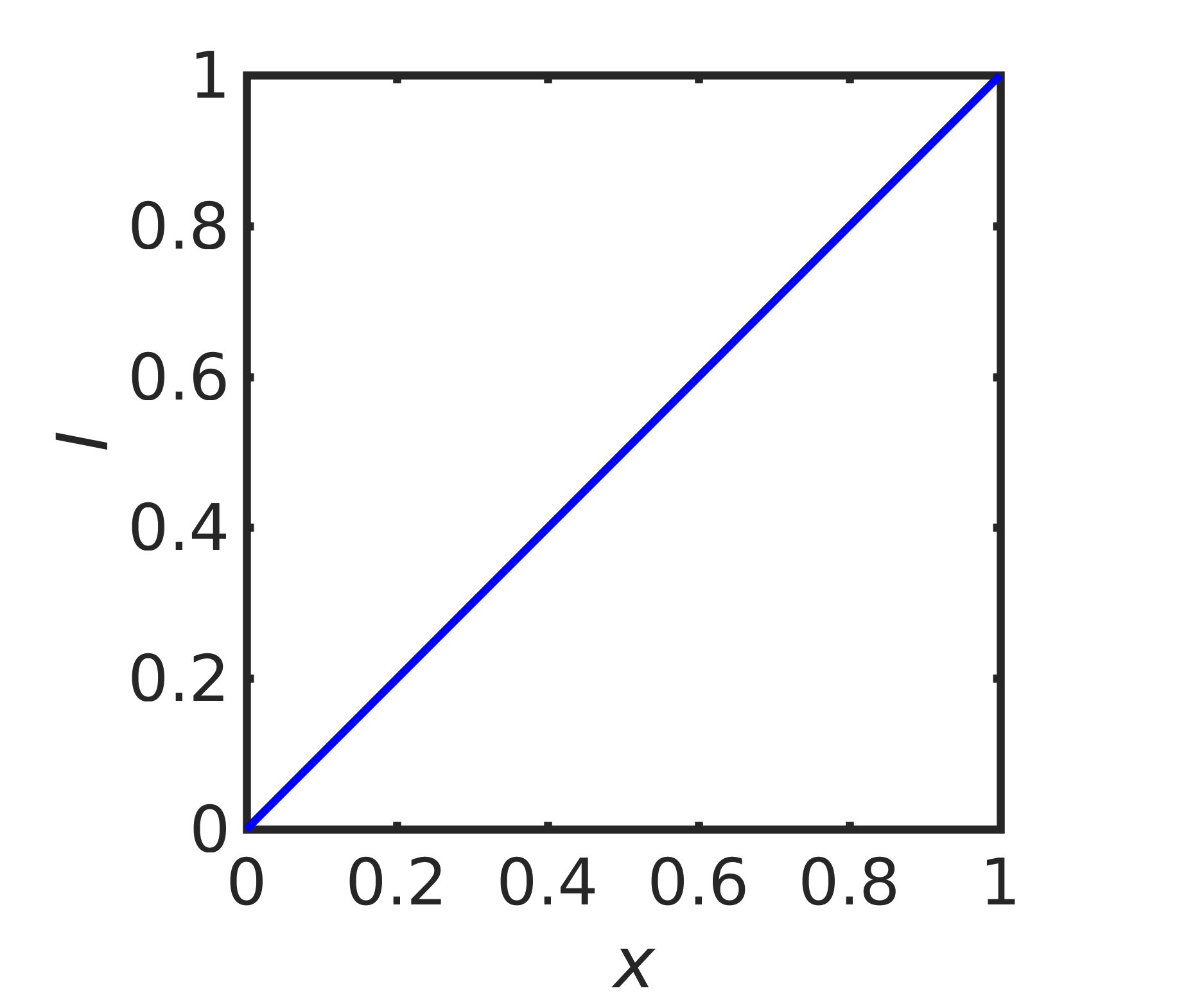}
     \includegraphics[width=0.49\linewidth]{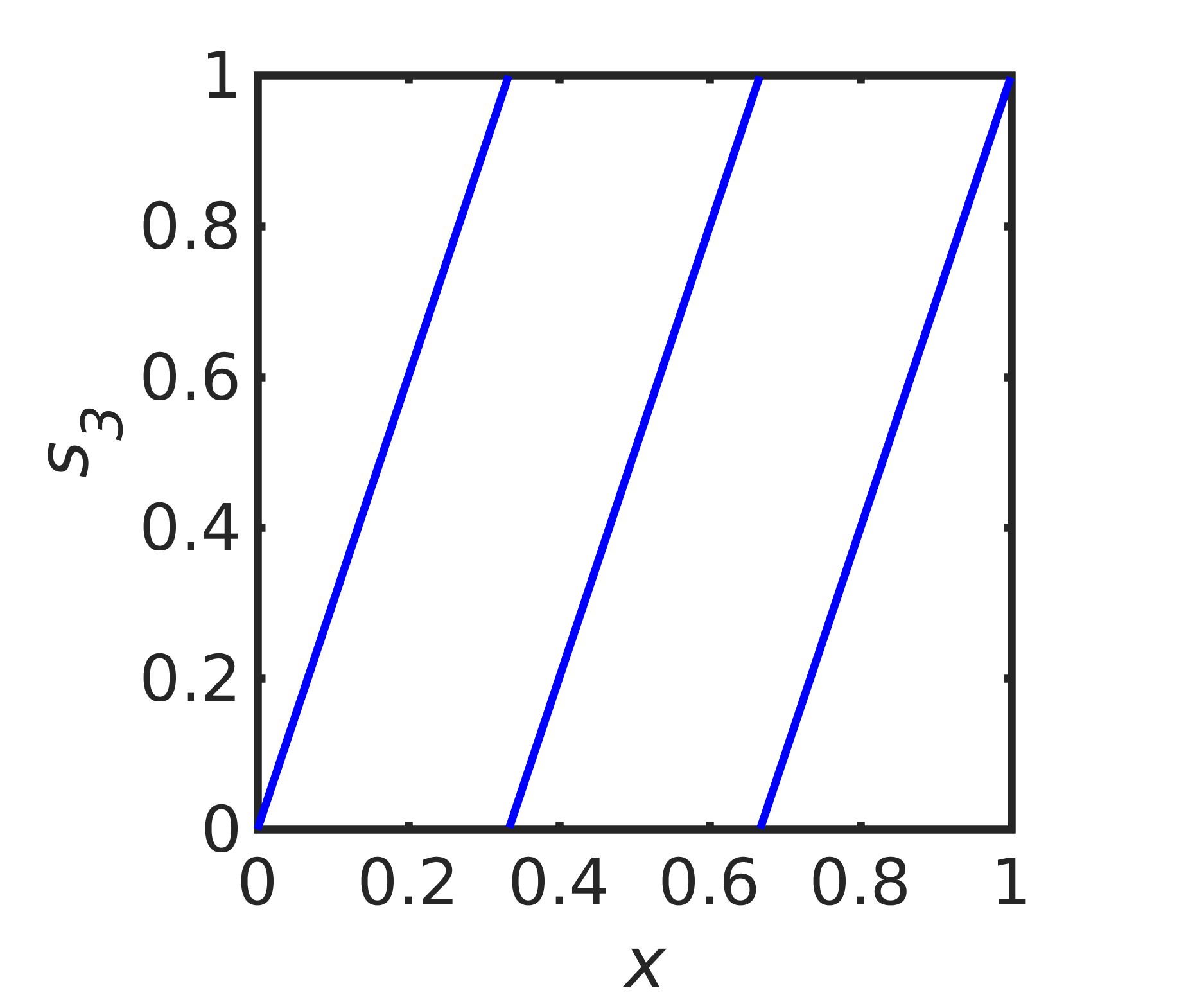}
     \includegraphics[width=0.49\linewidth]{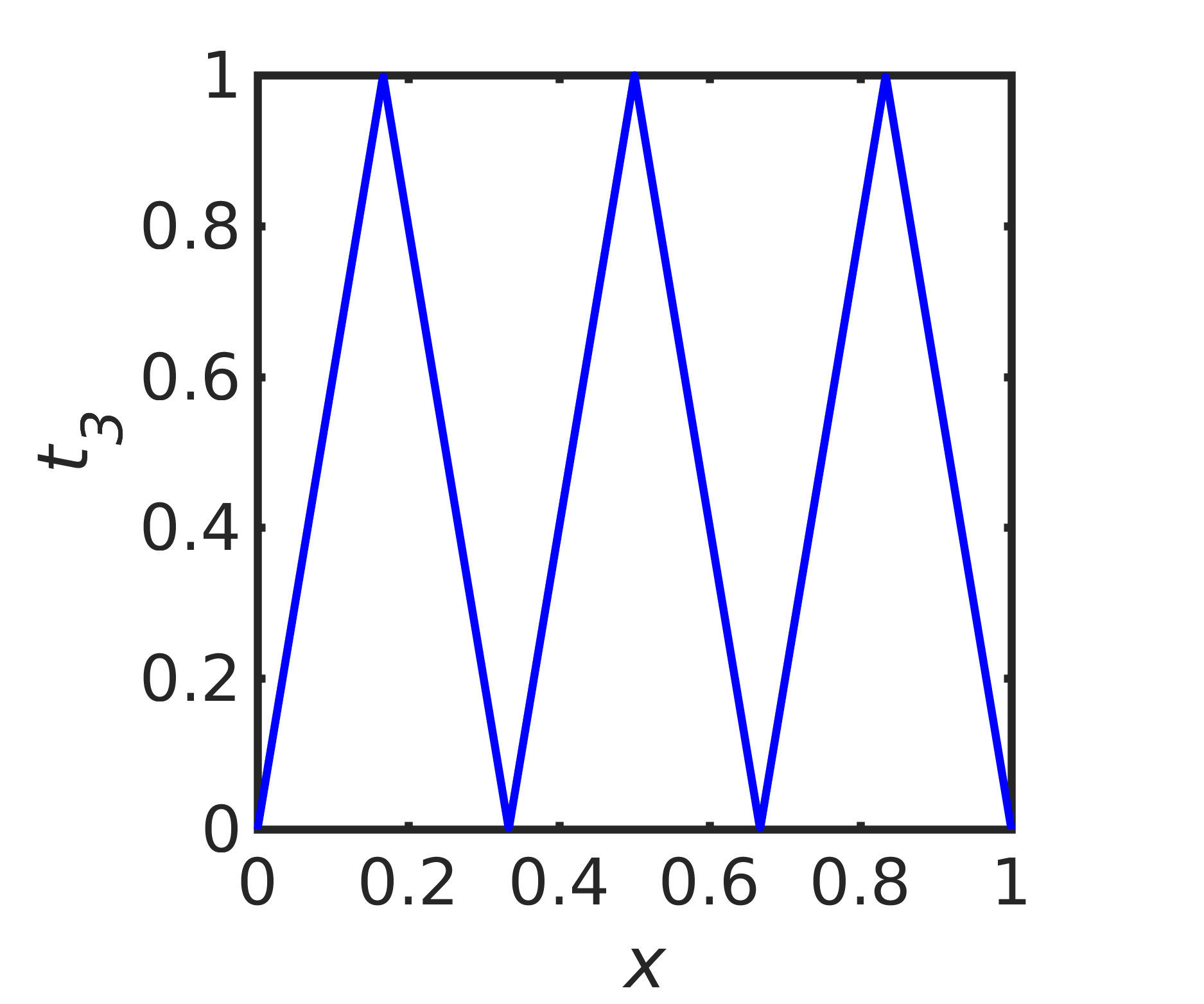}
     \includegraphics[width=0.49\linewidth]{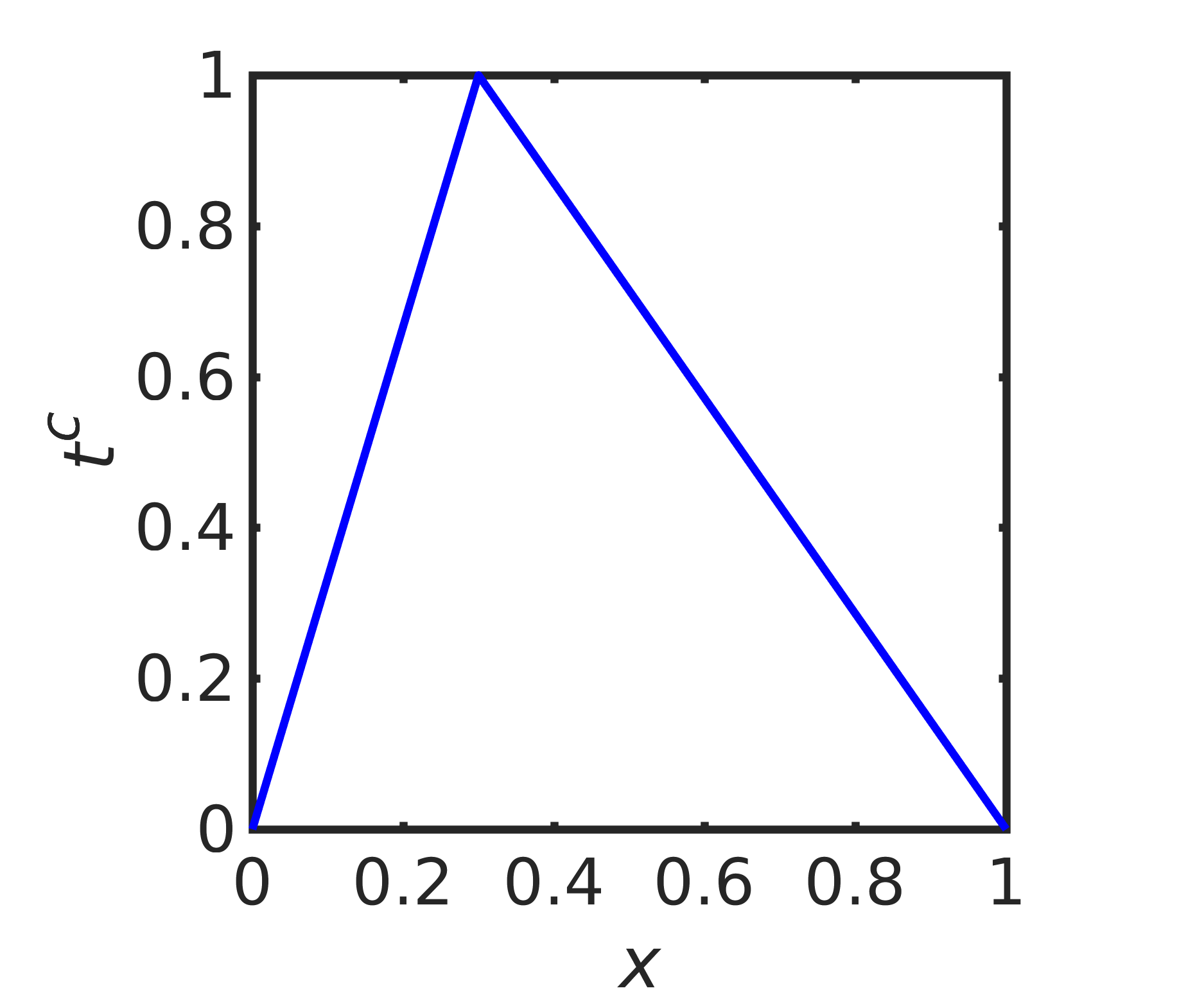}
     \caption{Four examples of uniform maps on $[0,1]$. Top-left: identity map. Top-right: $\ell$ periods of a sawtooth wave for $\ell=3$. Bottom-left: $\ell$ periods of a triangle wave for $\ell=3$. Bottom-right: asymmetric triangle for $c=0.3$.}
     \label{fig:uniform}
\end{figure}

We have already encountered three uniform maps on the interval $[0,1]$, namely the 
identity map $I(x)=x$ (Figure~\ref{fig:uniform} (top, left)) and the translation operator~\eqref{eq:Tc} (Figure~\ref{fig:Tct1} (left)), that have Lyapunov exponent $h=0$, and the triangle map (Figure~\ref{fig:Tct1} (right)) with Lyapunov exponent $h=\log 2$.

Some further elementary uniform maps on $[0,1]$, and associated Lyapunov exponents, are:
\begin{itemize}
 \item $\ell$ periods of a sawtooth function  on $[0,1]$ (Figure~\ref{fig:uniform}, top-right, for $l=3$ periods)~\footnote{The two period sawtooth map $s_2$ is also called the Bernoulli map, and its orbit $\mathcal{O}^+(x)$ is the dyadic transformation.}
       \begin{equation} s_\ell(x) = \ell*x-\lfloor \ell*x\rfloor ,\label{eq:saw}\end{equation}
       with Lyapunov exponent $h=\log\ell$,
 \item $\ell$ periods of a triangle function on $[0,1]$ (Figure~\ref{fig:uniform} (bottom, left) for $l=3$ periods)~\footnote{This is the `broken linear transformation' in~\cite{grossmann1977invariant} of order $p=2\ell$.} 
 \begin{equation} t_\ell(x) = 1-2\left|s_\ell(x)-1/2\right| .\label{eq:tri}\end{equation}
 with Lyapunov exponent $h=\log 2\ell$,
 \item and the asymmetric triangle, for $c\in (0,1)$ (Figure~\ref{fig:uniform} (bottom, right) for $c=0.3$)
 \begin{equation} t^c(x) = \begin{cases}
                    \displaystyle    \frac{x}{c}         & 0\le x \le c \\[1em]
                    \displaystyle    \frac{1-x}{1-c}     & c \le x \le 1
              \end{cases}
 \label{eq:asym}\end{equation}
 with Lyapunov exponent $0\leq h= -c\log c - (1-c)\log(1-c) \leq \log 2$.
\end{itemize}

Obviously, many more uniform maps are possible. Further examples can be formed by partitioning the domain and range of any uniform map, and then permuting the subintervals. Many existing `matrix-based' methods for constructing solutions to the IFPP, can be viewed as examples of such a partition-and-permute of an elementary uniform map~\cite{rogers2004synthesizing}. Uniform maps of other forms are developed in~\cite{huang2005characterizing} from two-segmental Lebesgue processes, producing uniform maps that are curiously non-linear.

\begin{lemma}
 \label{lem:comp}
 The composition of uniform maps is also a uniform map, i.e., if $U_1$ and $U_2$ are uniform maps then so is $U = U_1 \circ U_2$.
\end{lemma}

An example is $t_\ell$, that can be constructed as the composition $t_\ell = s_\ell \circ t_1$. 

We mentioned the numerical artifacts that can occur with finite-precision arithmetic, particularly when implementing maps on a binary computer and when the endpoints of the interval $X$  and constants in the maps have exact binary representations. We avoided these artifacts by composing the stated uniform map with the translation $T^c$ for $c=1/3\times10^{-9}$ that does not have finite binary representation~\footnote{Computation was performed in MatLab that implements IEEE Standard 754 for double-precision binary floating-point format.}. This small shift is indiscernible in the graphs of the maps.

\subsection{Ramp Distribution}
\label{sec:ramp}
\begin{figure} \centering
     \includegraphics[width=0.49\linewidth]{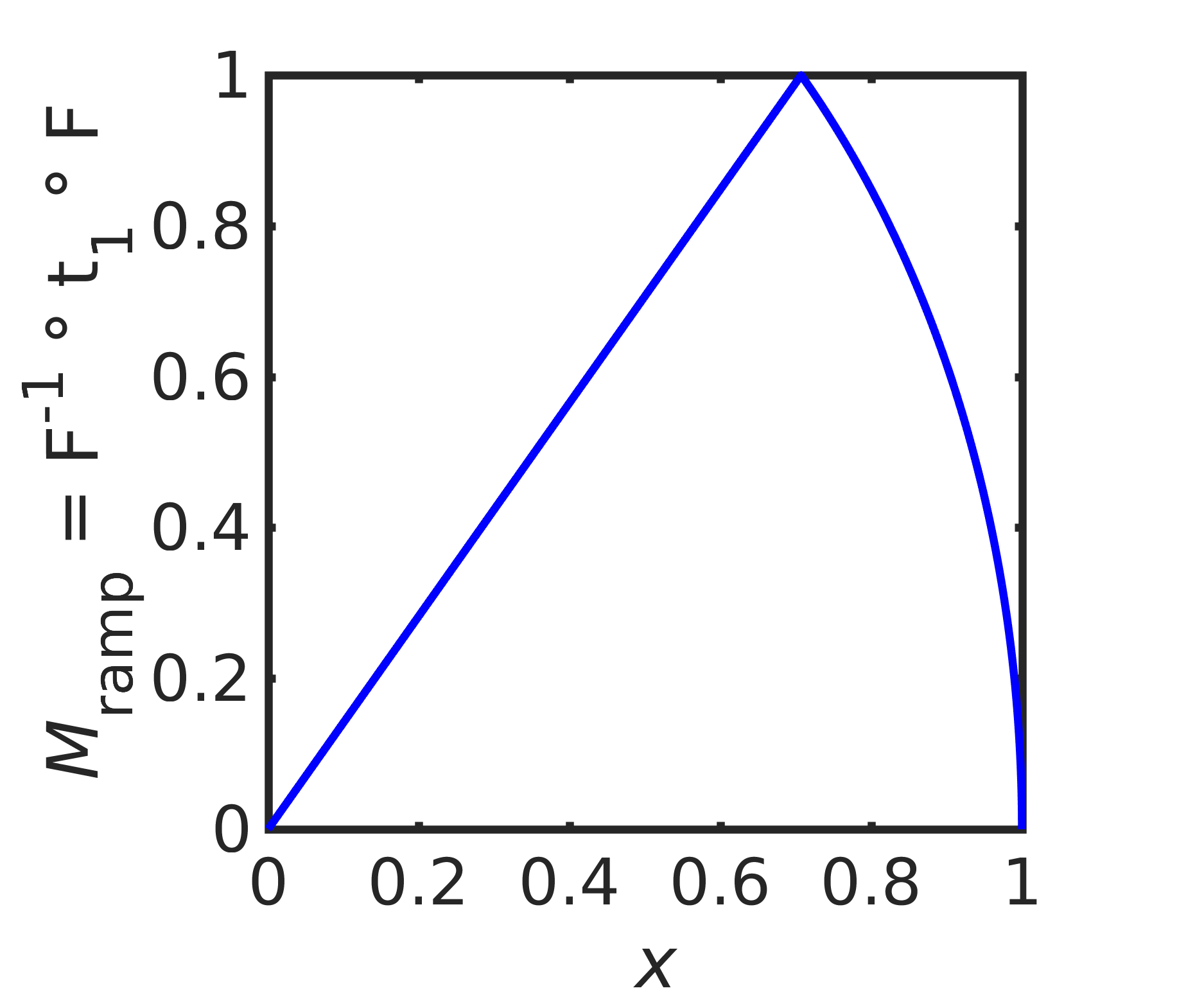}
     \includegraphics[width=0.49\linewidth]{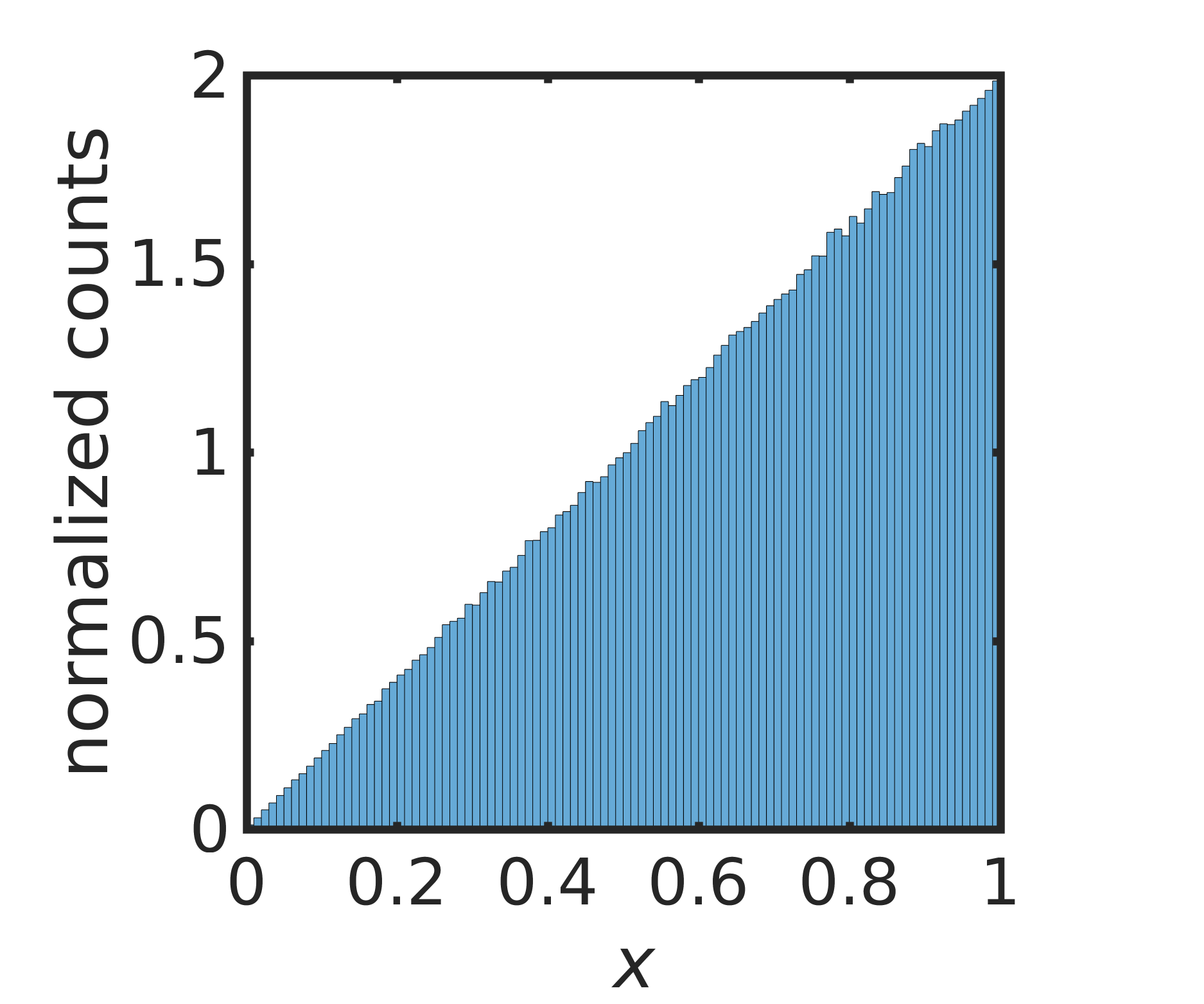}
     \caption{Iterative map $M_\text{ramp}$ in~\eqref{eq:Mramp} (left) and a normalized histogram of $1\times 10^6$ iterates  that approximates the equilibrium PDF (right).}
     \label{fig:Mramp}
\end{figure}

To give a concrete example of the solution in~\eqref{eq:sym} for a distribution without reflexive symmetry, we consider the ramp distribution with PDF
\begin{equation}
 \rho_\text{ramp}(x) = 2x
\end{equation}
that has CDF
\[ F_\text{ramp}(x) = x^2.
\]
We produce a unimodal, continuous map by choosing the uniform map $t_1$, as in~\eqref{eq:sym}, to give
\begin{equation}
  \label{eq:Mramp}
  M_\text{ramp} = 
  \begin{cases}
    \sqrt{2}x                  & 0\le x \le \frac{1}{\sqrt{2}} \\
    \sqrt{2}\sqrt{1-x^2}       & \frac{1}{\sqrt{2}} \le x \le 1
  \end{cases}
\end{equation} 
shown in Figure~\ref{fig:Mramp} (left). Figure~\ref{fig:Mtent} (right) shows a normalized histogram of $10^6$ iterates of $M_\text{ramp}$ starting at $x=0.3$, confirming that the orbit of $M_\text{ramp}$ converges to the desired ramp distribution, as guaranteed by Theorem~\ref{theo:fact}. The numerical implementation avoids finite-precision effects, as discussed earlier.
The estimated Lyapunov exponent for this map is $h\approx 1.040035$, which is greater than the Lyapunov exponent for the inducing triangular map $t_1$ which is $\log 2\approx  0.693147$.

\begin{figure} \centering
     \includegraphics[width=0.49\linewidth]{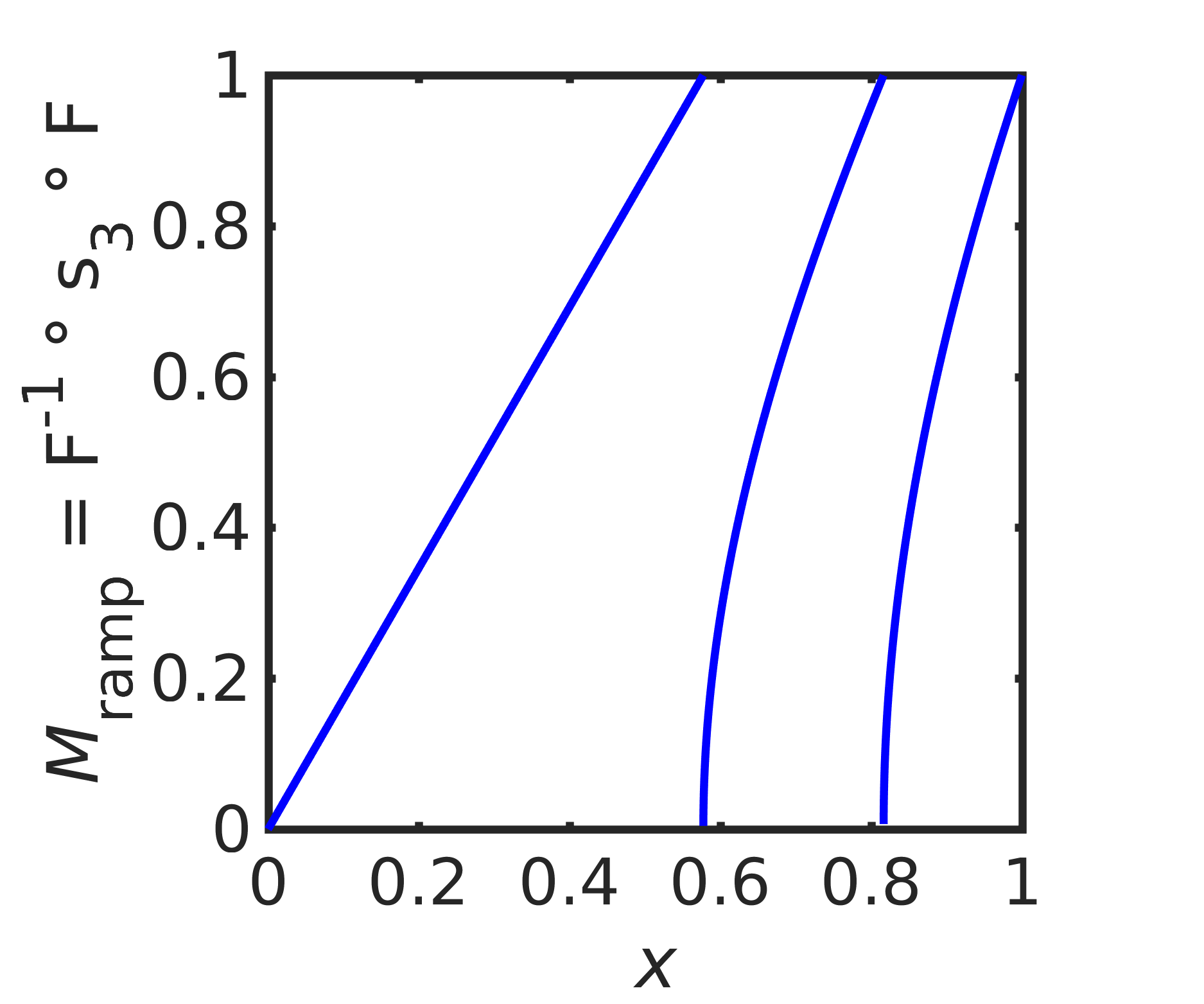}
     \includegraphics[width=0.49\linewidth]{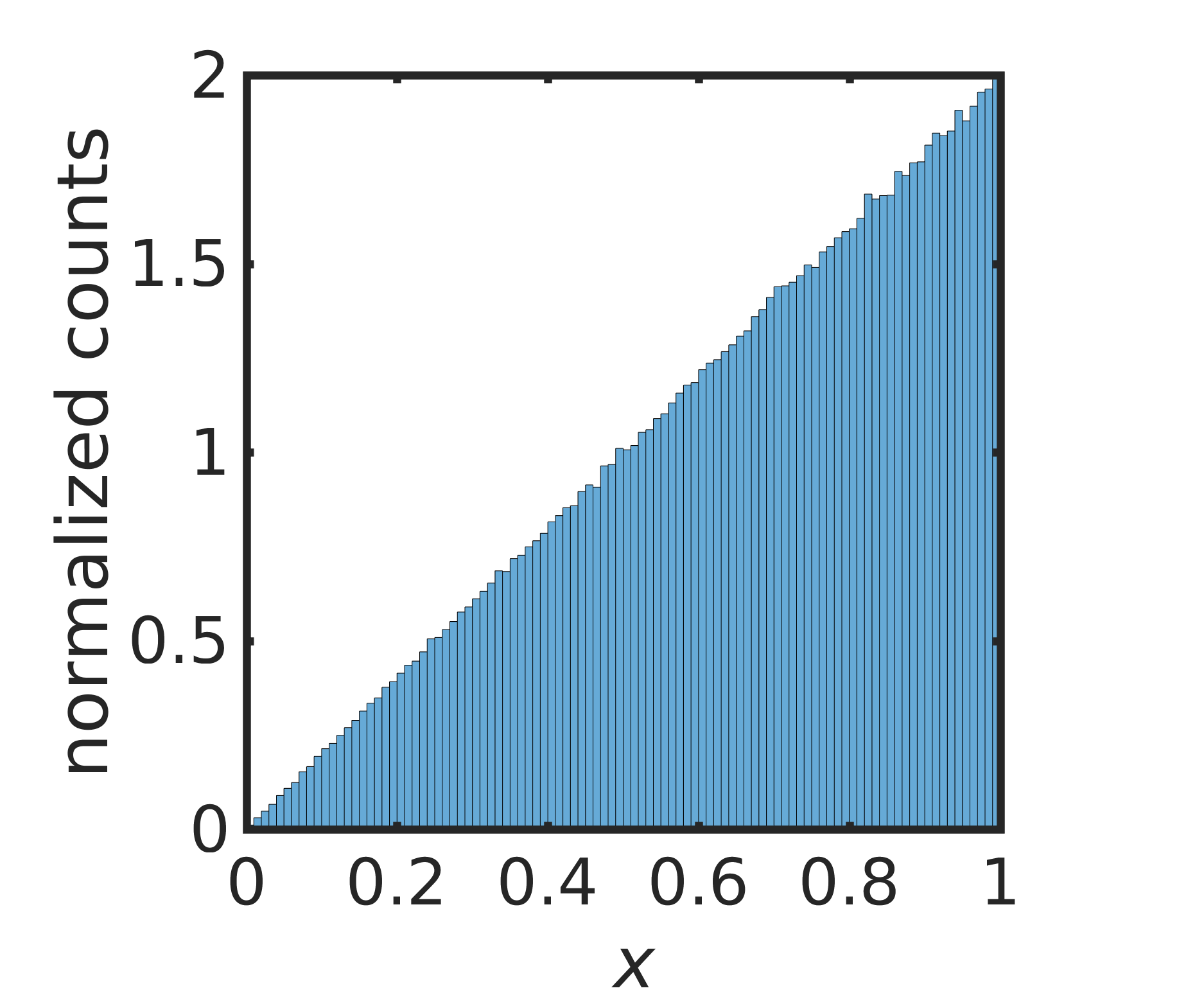}
     \caption{Iterative map $M_\text{ramp}$ in~\eqref{eq:Mramp} (left) and a normalized histogram of $1\times 10^6$ iterates  that approximates the equilibrium PDF (right).}
     \label{fig:MrampS}
\end{figure}

Using a different uniform map gives a different solution to the IFPP. For example, choosing $s_3$ gives the map $M=F^{-1}\circ s_3\circ F$ shown in Figure~\ref{fig:MrampS} (left). A normalized histogram over an orbit of $10^6$ iterations is shown in Figure~\ref{fig:MrampS} (right), confirming that this map is also ergodic for $\rho_\text{ramp}$. 
The estimated Lyapunov exponent for this map is $h\approx 1.098612$, which is the same numerical value as the Lyapunov exponent for the sawtooth map $s_3$  which is $\log 3\approx  1.098612$. Comparing with the result for the map induced by $t_1$, this shows that the Lyapunov exponent of a map is not generally equal to the Lyapunov exponent of the inducing uniform map.

\subsection{The Logistic Map and Alternatives}
\label{sec:altlog}
The logistic map, mentioned in the introduction, is 
\begin{equation} 
  M_\text{log}(x) = 4x(1-x).
  \label{eq:logisticmap}
\end{equation}
The equilibrium distribution of this map
\begin{equation} \rho_\text{log}(x)=\frac{1}{\pi\sqrt{x(1-x)}}, \label{eq:logisticdist} \end{equation} 
can be easily verified by substituting into the FP equation~\eqref{eq:FP1dim}. The CDF of $\rho_\text{log}(x)$ is
\begin{equation}
 F_\text{log}(x) = \int_0^x \rho_\text{log}(x') dx' = \frac{2}{\pi}\arcsin(\sqrt{x}),
 \label{eq:logCDF}
\end{equation} 
and the IDF is
\begin{equation} 
  F_\text{log}^{-1}(x) = \sin^2(\frac{\pi x}{2}) = \frac{1}{2}(1-\cos(\pi x)). 
  \label{eq:logIDF}
\end{equation} 

\begin{figure} \centering
     \includegraphics[width=0.49\linewidth]{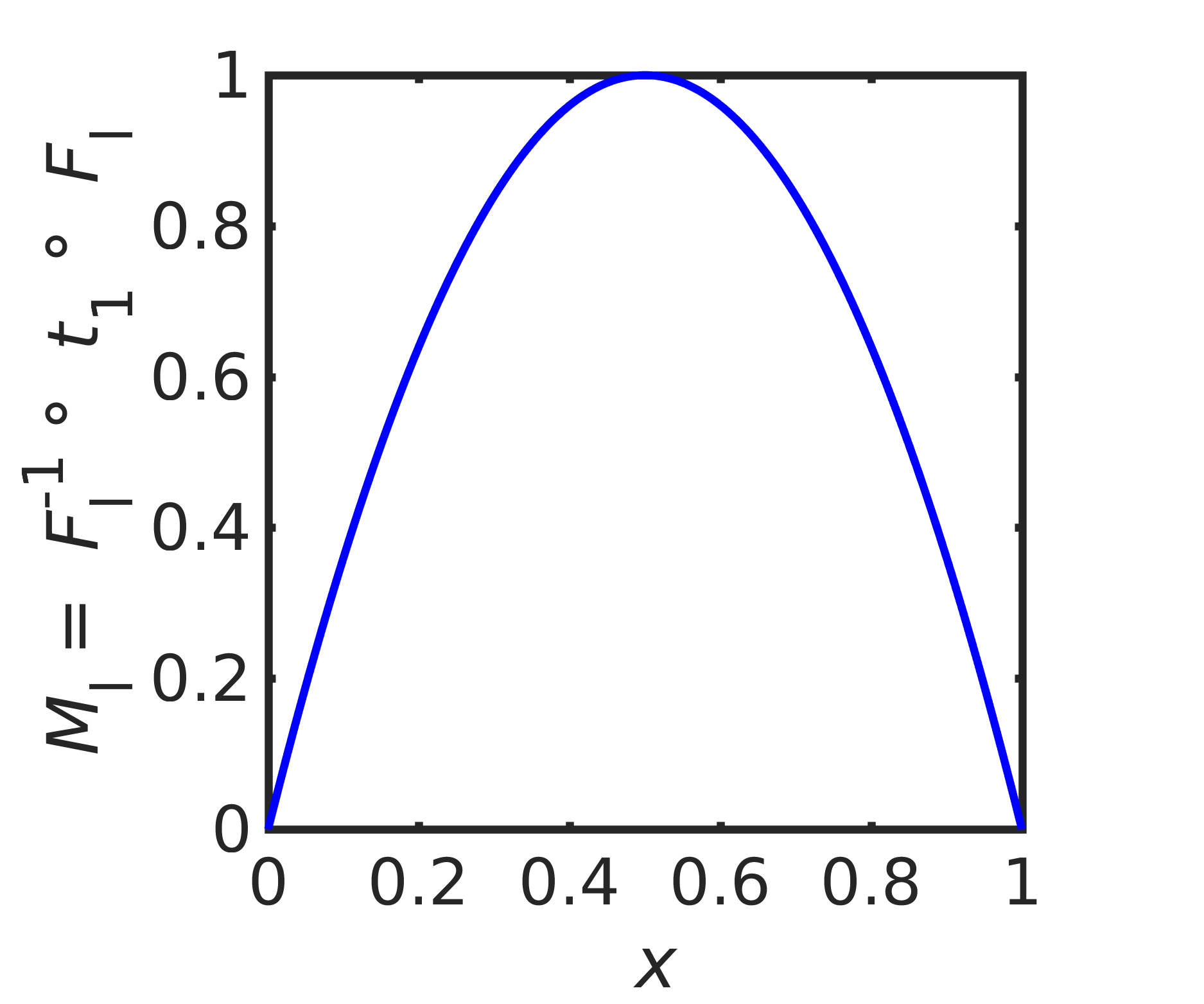}
     \includegraphics[width=0.49\linewidth]{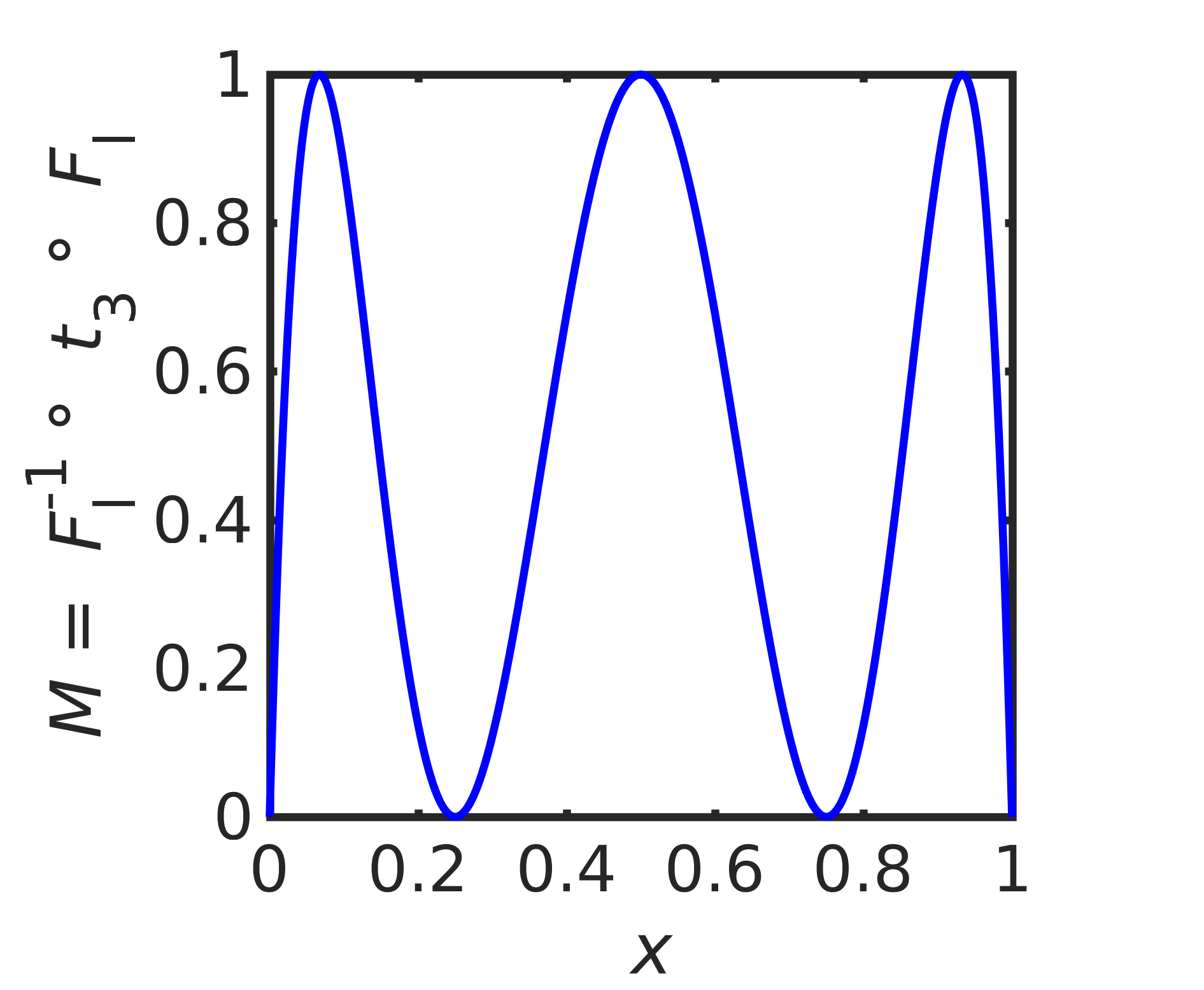}
     \caption{Logistic map, that equals $M_\text{log}=F_\text{log}^{-1}\circ t_1 \circ F_\text{log}$ (left), and the map $M=F_\text{log}^{-1}\circ t_3 \circ F_\text{log}$ that has the same equilibrium distribution (right).}
     \label{fig:logmaps}
\end{figure}

The logistic map~\eqref{eq:logisticmap} is induced by the factorization~\eqref{eq:fact} by choosing the triangle map $t_1$ as uniform map, i.e., substituting the CDF~\eqref{eq:logCDF} and IDF~\eqref{eq:logIDF} into~\eqref{eq:sym}; see Figure~\ref{fig:logmaps} (left). Equivalently, one may note that the logistic map~\eqref{eq:logisticmap} is transformed into the triangle map $t_1$ by the change of variables $z=F^{-1}(x)$; in the language of~\cite{grossmann1977invariant}, $M_\text{log}$ and $t_1$ are \emph{conjugate} dynamical laws. 

Other iterative maps which preserves the same equilibrium distribution~\eqref{eq:logisticdist} can be constructed by choosing another uniform map, such as $\ell$ periods of a triangle function~\eqref{eq:tri}. This gives the iterative maps
\begin{equation}
 M_\ell=F_\text{log}^{-1}\circ t_\ell \circ F_\text{log} = \sin^2(2\ell \arcsin(\sqrt{x})), \ell\geq 1,
 \label{eq:logalt}
\end{equation}
that coincide with the $n^\text{th}$ power of the logistic map~\eqref{eq:logisticmap} for $\ell=2^{n-1}$. 
Figure~\ref{fig:logmaps} (right) shows the map which preserves the same equilibrium distribution as the logistic map but induced by the uniform map $t_3$. Since $3$ is not of the form $2^{n-1}$, this map is not simply a power of the logistic map.

The theoretical value of the Lyapunov exponent of the map in~\eqref{eq:logalt} is $\log 2\ell$, using~\eqref{eq:LyapExpect}. 
Table~\ref{tab:1} gives the theoretical values of the Lyapunov exponent and experimentally calculated values using $10000$ iterations, as in~\eqref{eq:Lyap}, for some values of $\ell$.
\begin{table}[h]
\begin{tabular}{|l|l|l|}
\hline
 $\ell$   & $h_\text{e}$ & $h_\text{t}$ \\ \hline
 $1$      & $0.692819$   & $0.693147$   \\ \hline
 $2$      & $1.386284$   & $1.386294$   \\ \hline
 $4$      & $2.079430$   & $2.079442$   \\ \hline
 $2^{24}$ & $17.328594$  & $17.328680$  \\ \hline
 $2^{39}$ & $27.725713$  & $27.725887$  \\ \hline
\end{tabular}
\caption{Experimental $h_\text{e}$ and theoretical $h_\text{t}$ Lyapunov exponents for the maps in~\eqref{fig:logmaps} for a range of $\ell$, given to $6$ decimal places. \label{tab:1}}
\end{table}

\section{Two Examples in Two Dimensions}
\label{sec:twodim}
\subsection{Uniform Maps on $[0,1]^2$}
Two well-known examples of uniform maps in the two-dimensional unit square, $X=[0,1]^2$, are the baker's map
\begin{equation}
  U_\text{b}(x_1,x_2) = \left( 2x_1\mod 1, \frac{x_2+u\left(x_1-\frac{1}{2}\right)}{2} \right) ,
  \label{eq:bakers}
\end{equation}
where $u$ is the unit step function, and the Arnold cat map
\begin{equation} 
  U_\text{A}(x_1,x_2) = \left( (2x_1+x_2)\mod 1, (x_1+x_2)\mod 1 \right).
  \label{eq:arnold}
\end{equation}
Other uniform maps in $d>1$ dimensions may be formed by $1$-dimensional uniform maps acting on each coordinate, giving the coordinate-wise uniform map
\begin{equation} 
  U(x) = \left(U_1(x_1), U_2(x_2),\ldots, U_n(x_d)\right) 
  \label{eq:compunif}
\end{equation}
where $U_i(x)$, $i=1,2,\ldots,d$, are uniform maps in one dimension. We will use the baker's map~\eqref{eq:bakers} and a coordinate-wise uniform map in the $2$-dimensional examples that follow.

\subsection{Checker-Board Distribution}

This example demonstrates construction of a map in two-dimensions that targets the checker board distribution, shown in Figure~\ref{fig:checker_baker} (bottom-left) and Figure~\ref{fig:checker_twintent} (bottom-left), using the factorization~\eqref{eq:fact}. 

\begin{figure} \centering
     \includegraphics[width=1\linewidth]{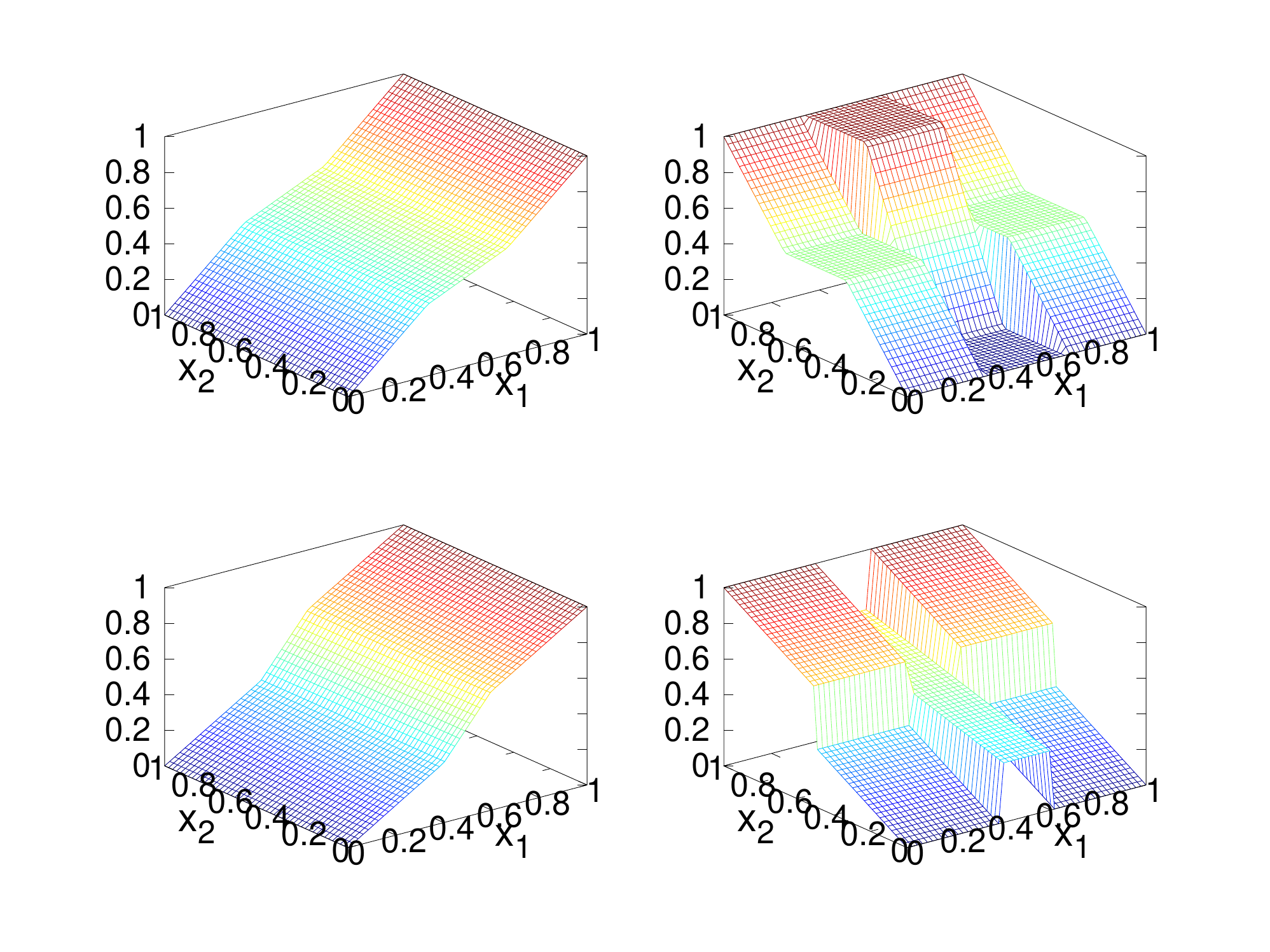}
     \caption{Plots of the Rosenblatt transformation $R$ and its inverse $R^{-1}$ for the checker-board distribution. Top row shows the components of $R$: first component (left) and second component (right). Bottom row shows the components of $R^{-1}$: first component (left) and second component (right).}
     \label{fig:checker_FI}
\end{figure}

\begin{figure} \centering
     \includegraphics[width=1\linewidth]{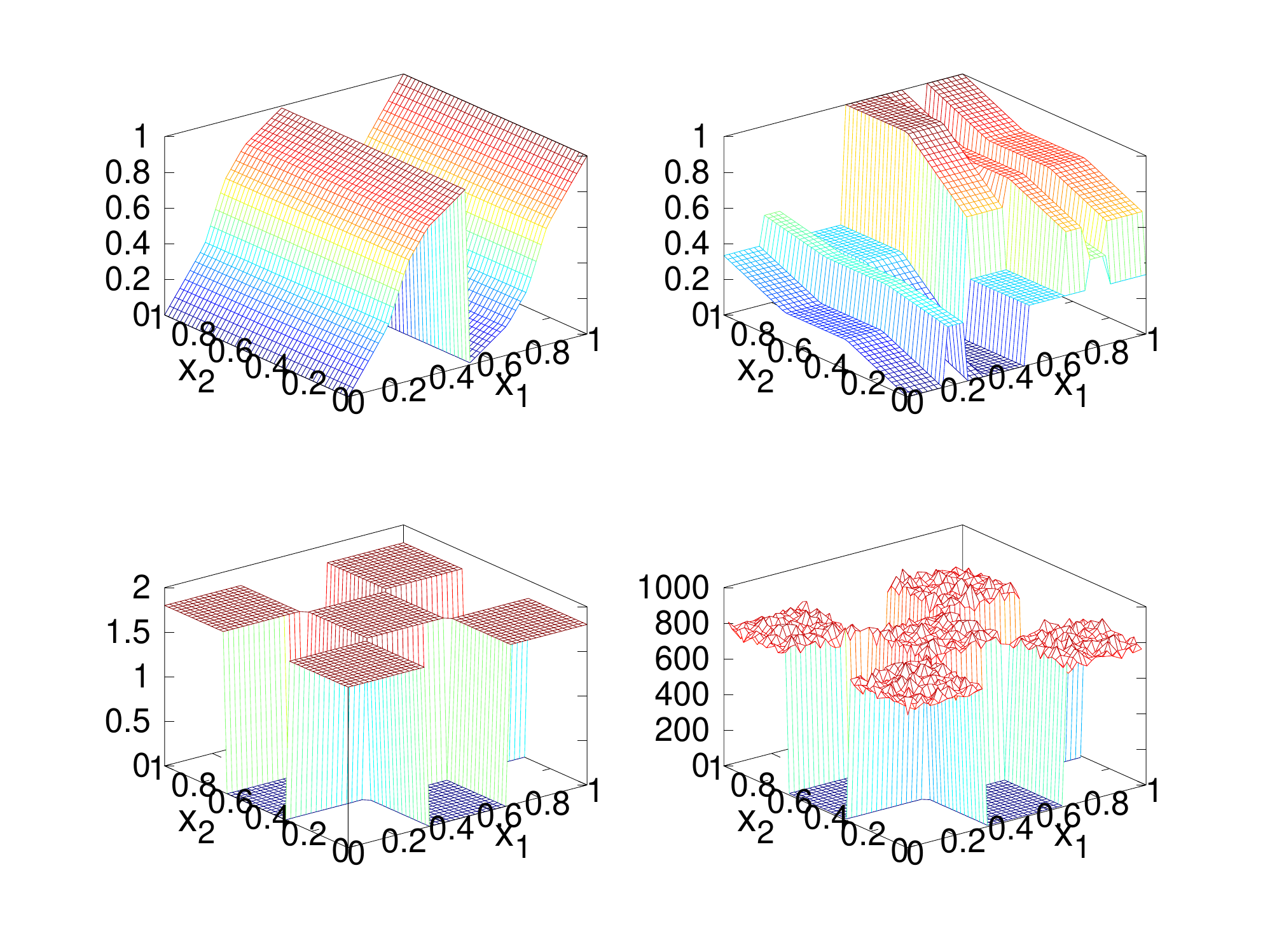}
     \caption{Iterated function constructed with $U$ being the baker's transformation, a histogram of iterates, and the checker board distribution. Shown is: the $x_1$ part of the constructed map (top-left), the $x_2$ part of the constructed map (top-right), the checker board distribution (bottom-left), and a histogram of iterates of the constructed map (bottom-right).}
     \label{fig:checker_baker}
\end{figure}

\begin{figure} \centering
     \includegraphics[width=1\linewidth]{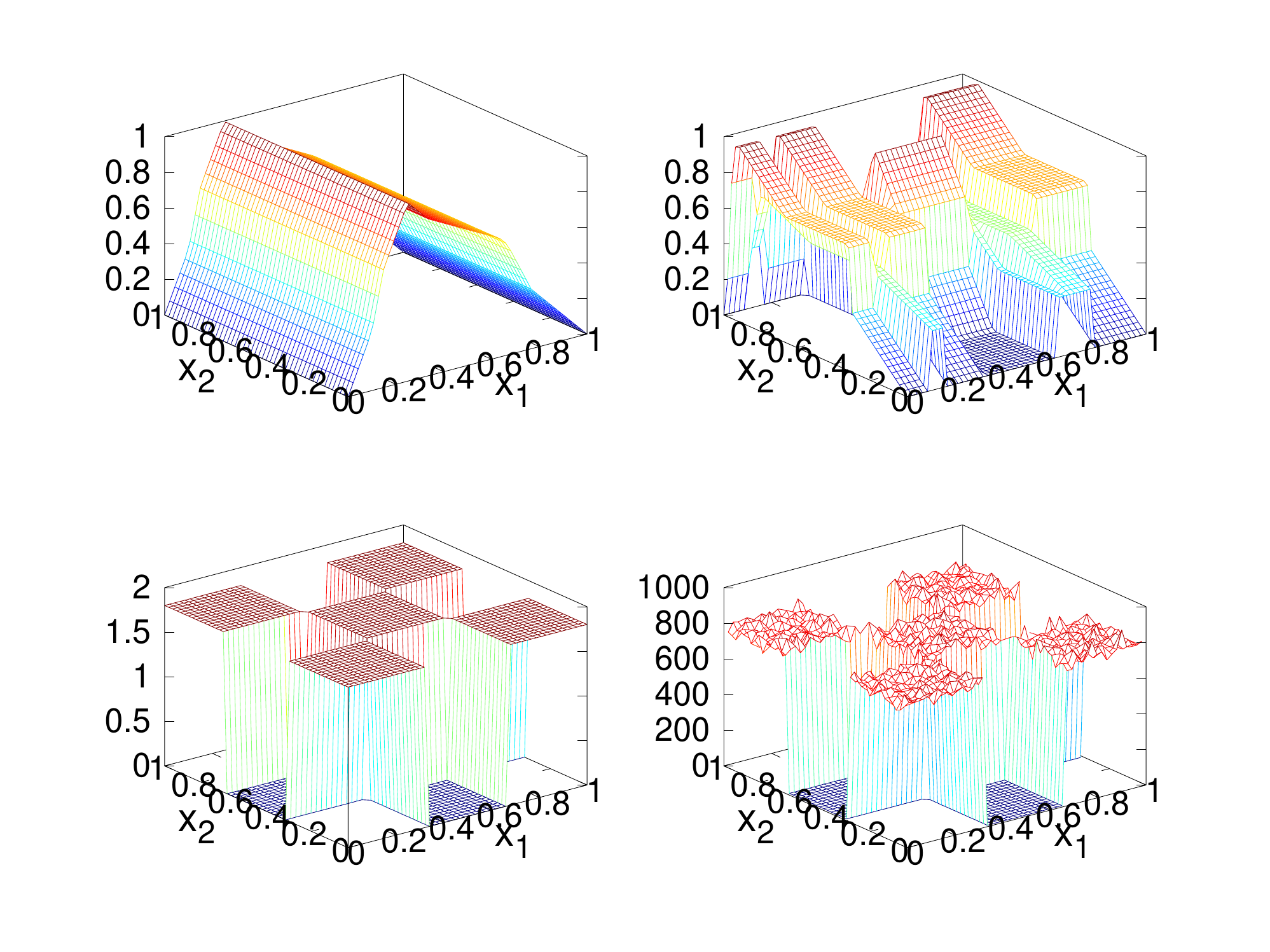}
     \caption{Same as Figure~\ref{fig:checker_baker} except $M$ is constructed with $U$ being component-wise asymmetric triangular maps.}
     \label{fig:checker_twintent}
\end{figure}

The first step in constructing a solution to the IFPP for this distribution is to construct the forward and inverse Rosenblatt transformations, which requires the marginal density functions~\eqref{eq:mdf} that may be evaluated analytically in this case. A plot of the two components of the functions $R$ and $R^{-1}$ is shown in Figure~\ref{fig:checker_FI}.

We construct two solutions to the IFPP, each induced by choosing a particular uniform map: the first is the baker's map~\eqref{eq:bakers}; the second is a component-wise uniform map~\eqref{eq:compunif} with an asymmetric triangle map~\eqref{eq:asym} acting on each component,
 \begin{equation} U(x_1,x_2) = \left( U_1(x_1), U_2(x_2) \right) \end{equation}
where $U_1 =t^c$ for $c=0.3$ and $U_2=t^c$ with $c=0.9$.

Figure~\ref{fig:checker_baker} (top row) shows the two components of the map induced by the baker's map, the checker-board distribution (bottom-left), and a histogram of $10^6$ iterates (bottom-right) showing that the map does indeed converge in distribution to the desired distribution.

Figure~\ref{fig:checker_twintent} (top row) shows the two components of the map constructed using the two component-wise asymmetric triangular maps, the checker-board distribution (bottom-left), and a histogram of $10^6$ iterates (bottom-right) showing that the map also converges in distribution to the desired distribution, and hence is also a solution to the IFPP.

\subsection{A Numerical Construction}
\label{sec:num}
\begin{figure} \centering
   \includegraphics[width=0.4\linewidth]{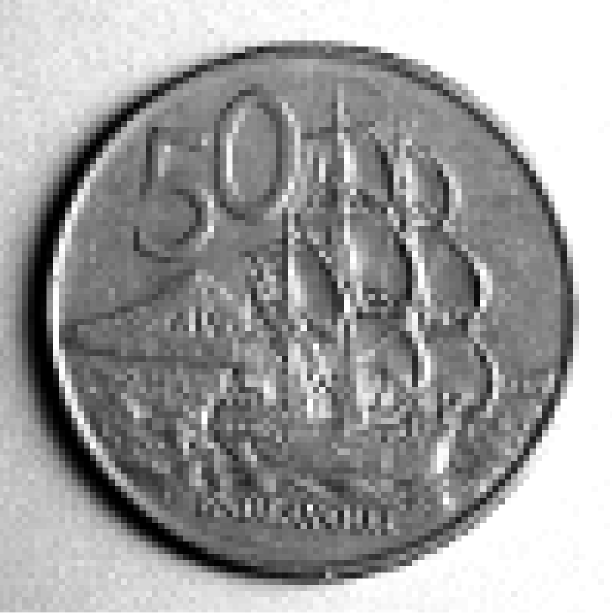}
   \hspace*{0.15\linewidth}
   \includegraphics[width=0.4\linewidth]{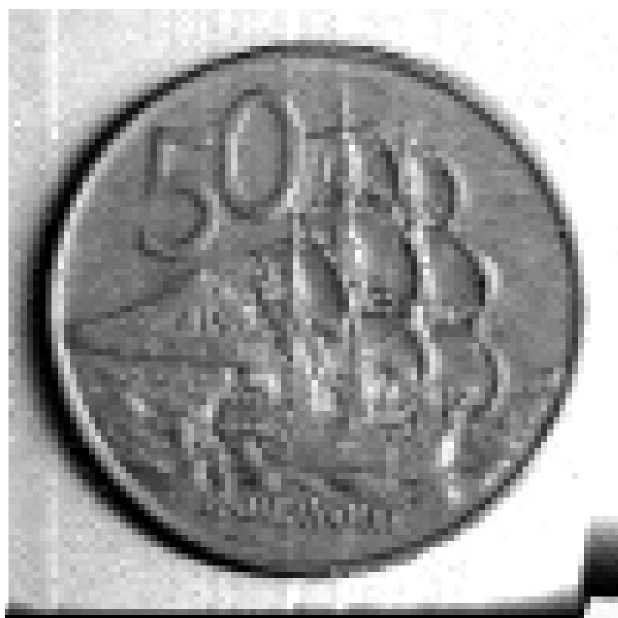}
     \caption{A normalized greyscale image of a coin used as the target distribution, and a normalized histogram of iterations of the map targeting this distribution. Original image (left), and normalized histogram of $10^6$ iterations (right).}
     \label{fig:coins}
\end{figure}

Numerical implementation of the factorized solution~\eqref{eq:fact} is not difficult in few dimensions. In this section we present an example of numerical implementation using a normalized greyscale image of a coin~\footnote{A pre-2006 New Zealand $50$ cent coin.}, piecewise constant over pixels, as the target distribution; see Figure~\ref{fig:coins} (left). The marginal distributions~\eqref{eq:mdf} are evaluated as linear interpolation of cumulative sums over pixel values, and hence the CDF and then forward and inverse Rosenblatt transformations follow as in section~\ref{sec:rosen}. The uniform map was produced as component-wise univariate translation maps, specifically 
\[ U(x_1,x_2) = \left( U_1(x_1), U_2(x_2) \right) \]
where $U_1 =T^c$ for $c=0.6$ and $U_2=T^c$ with $c=0.2$.
The resulting map is given by Eq.~\eqref{eq:fact}.

Figure~\ref{fig:coins} (right) shows a normalized histogram, binned to pixels, of $10^6$ iterations of this map. As can be seen, the estimated PDF from the orbit of this map does reproduce the image of the coin. However, there are also obvious artifacts near the edge of the image showing that mixing could be better. We conjecture that a chaotic uniform map would produce better mixing and fewer numerical artifacts.

\section{Summary and Discussion}
\label{sec:sad}
We have shown how the solution of the IFPP, of finding an iterative map with a given invariant distribution, can be constructed from uniform maps through the factorization established in Theorem~\ref{theo:fact},
 \[ M = R^{-1} \circ U \circ R \]
where $R$ denotes the Rosenblatt transformation that has Jacobian determinant equal to density function of the invariant distribution. In one-dimension, $R$ is exactly the CDF of the given distribution, so the factorization generalizes existing one-dimensional solutions to the setting of arbitrary multi-variate distributions. The factorization also shows the relationship between arbitrary iterative maps and uniform maps, i.e., given a Rosenblatt transformation the solution of the IFPP is equivalent to the choice of a uniform map that has $\operatorname{Unif}([0,1]^d)$ as invariant distribution.

We find the factorization~\eqref{eq:fact} appealing as it shows that solution of the IFPP for arbitrary distributions, and in multi dimensions, is reduced to two standard and well-studied problems, i.e., constructing the Rosenblatt transformation (or CDF in one dimension) and designing a uniform map. It is therefore surprising, to us, that the factorization~\eqref{eq:fact}, and more generally the Rosenblatt transformation, appears to not be widely used in the study of chaotic iterated functions and the IFPP.  Grossmann and Thomae~\cite{grossmann1977invariant}, in one of the earliest studies of the IFPP, essentially derived the factorization~\eqref{eq:fact} by introducing conjugate maps and establishing the relation (in their notation)  that $\rho^*(x)=\mathrm{d} h^{-1}(x)/\mathrm{d} x$ where $\rho^*$ is the invariant distribution and $h$ is the conjugating function; see~\cite[Figure 3]{grossmann1977invariant}. It is a small step to identify that $h$ is the IDF, generalized in multi-dimensions by the inverse Rosenblatt transformation. However, the connection was not made in~\cite{grossmann1977invariant}, despite the Rosenblatt transformation having been already known, in statistics, for some decades~\cite{rosenblatt-1952}. 

We constructed solutions to the IFPP for distributions with special reflexive symmetry structure, and then with no special structure, by constructing the Rosenblatt transformation, and its inverse, for some examples in one and two dimensions. For simple distributions with analytic form the Rosenblatt transformation may be constructed analytically, while numerically-defined distributions required calculation of the marginal distributions~\eqref{eq:mdf} using numerical techniques. 

Although this factorization and construction is applicable to high-dimensional problems, the main difficulty is obtaining all necessary marginal densities, which requires the high-dimensional integral over $x_{k+1}\ldots x_d$ in \eqref{eq:mdf}. In general, this calculation can be extremely costly. Even a simple discretization of the PDF $\rho$, or of the argument of the marginal densities~\eqref{eq:mdf}, leads to exponential cost with dimension.

To overcome this cost, Dolgov \emph{et al.}~\cite{dolgov2020approximation} precomputed an approximation of $\rho(x_1,\ldots,x_d)$ in a compressed tensor train representation that allows fast computation of integrals in~\eqref{eq:mdf}, and subsequent simulation of the inverse Rosenblatt transformation $R^{-1}$ from the conditionals in \eqref{eq:mdf}, and showed that computational cost scales \emph{linearly} with dimension $d$. Practical examples presented in~\cite{dolgov2020approximation}, in dimension $d\leq 32$, demonstrate that operation by the forward and inverse Rosenblatt transformations is computationally feasible for multivariate problems with no special structure.  

Finding a solution of the IFPP with desired properties is reduced to a standard problem of designing a uniform map on $[0,1]^d$, for which there are many existing efficient options. For example, standard computational uniform random number generators, that produce pseudo-random sequences of numbers, are one such existing uniform map, as are the quasi-Monte Carlo rules that we mentioned earlier~\cite{dick2013high}. These induce pseudo-random and quasi-Monte Carlo sequences, respectively, on the space $X$ via the inverse Rosenblatt transformation $R^{-1}$~\cite{gentle2003random}.  Both these schemes were demonstrated in practical high-dimensional settings in~\cite{dolgov2020approximation}.

The RHS of Eq.~\eqref{eq:commute} in $d=1$ dimension is exactly the standard computational route for implementing inverse cumulative transformation sampling from $\rho$, since computational uniform pseudo-random number generators perform a deterministic iteration on $[0,1]$ to implement a uniform map~\cite{gentle2003random}. For $d>1$, the RHS of Eq.~\eqref{eq:commute} is the conditional distribution method that  generalizes the inverse cumulative transformation method, as mentioned above~\cite{devroye-rvgen-1986,johnson1987multivariate,hormann-rvgen-2004,dolgov2020approximation}. Hence Lemma~\ref{lem:commute} shows that standard computational implementation of both the inverse cumulative transformation method in $d=1$ and the conditional distribution method in $d>1$ is equivalent to implementing a solution to the IFFP. In this sense, computational inverse cumulative transformation sampling from $\rho$ can be viewed as the prototype for all iterative maps that target the distribution $\rho$, with each ergodic sequence corresponding to a particular choice of uniform map.

We mentioned that the Rosenblatt transformation associated with a given distribution $\rho$ is not unique. Actually, any two Rosenblatt transformations for $\rho$ are related by a uniform map, as shown in the following Lemma.
\begin{lemma}
 If $R_1$ is a Rosenblatt transformation for $\rho$ then $R_2$ is a Rosenblatt transformation for $\rho$ if and only if
 \[ R_2 = U\circ R_1\] 
 for some uniform map $U$.
\end{lemma}
\noindent \textbf{Proof:}
($\Rightarrow$) Since $R_1$ and $R_2$ are Rosenblatt transformations for $\rho$, then $U=R_2\circ R_1^{-1}$ is a uniform map and $R_2 = U\circ R_1$. ($\Leftarrow$) If $R_2 = U\circ R_1$ then if $x\sim\rho$, $R_1(x)\sim \operatorname{Unif}([0,1]^d)$ and $U\circ R_1(x)\sim \operatorname{Unif}([0,1]^d)$ so $R_2$ is a Rosenblatt transformation. 
\quad\qed

Hence, any Rosenblatt transformation $R$ may be written as $R = U\circ R_0$ for some uniform map $U$ and a fixed Rosenblatt transformation $R_0$.

The Rosenblatt transformations that map any distribution to the uniform distribution on the hypercube may also be used to understand mappings between spaces that are designed to transform one distribution to another, such as the `transport maps' developed in~\cite{parno2018transport}. Consider distributions $\rho_A$ and $\rho_B$, with Rosenblatt transformations $R_A$ and $R_B$, respectively, that may be related as in the following diagram. 
\[
 \begin{tikzcd}
   \rho_A \arrow[r, shift left=1.0ex, "\displaystyle R_A"] 
   & \operatorname{Unif}([0,1]^d) \arrow[l, shift left=1.0ex, "\displaystyle R_A^{-1}"] 
   \arrow[r,  shift left=1.0ex, "\displaystyle R_B^{-1}" ] 
   \arrow[loop above]{}{\displaystyle U}
   & \rho_B  \arrow[l, shift left=1.0ex, "\displaystyle  R_B" ] 
 \end{tikzcd}
\]
The diagram suggests a proof of the following lemma, that generalizes the factorization Theorem~\ref{theo:fact}.
\begin{lemma}
 A map $M$  satisfies $x\sim \rho_A \Rightarrow M(x)\sim\rho_B$ iff it can be written as $M= R_B^{-1}\circ U \circ R_A$, where $R_A$ and $R_B$ are Rosenblatt transformations for $\rho_A$ and $\rho_B$, respectively, and $U$ is a uniform map.
\end{lemma}

Hence, for given Rosenblatt transformations, the choice of a map, that maps samples from $\rho_A$ to samples from $\rho_B$, is equivalent to the choice of a uniform map. Alternatively, if a fixed uniform map is selected, such as the identity map, the choice of map $M$ is completely equivalent to the choice of Rosenblatt transformations. This factorization also shows that the equivalence class of conjugate maps, noted in~\cite{grossmann1977invariant}, for each dimension $d$, is generated by the uniform maps, and each member of the equivalence class contains maps that target each distribution, when the associated Rosenblatt transformation satisfies the mild conditions to be a conjugating function as defined in~\cite{grossmann1977invariant}. 


\bibliographystyle{unsrt}
\bibliography{IFPP}

\begin{thebibliography}{10}

\bibitem{Note1}
The density function is with respect to some underlying measure, typically
  Lebesgue. Throughout this paper we will use the same symbol for the
  distribution and associated PDF, with meaning taken from context.

\bibitem{dorfman1999introduction}
J.R. Dorfman.
\newblock {\em Cambridge Lecture Notes in Physics: An introduction to chaos in
  nonequilibrium statistical mechanics}, volume~14.
\newblock Cambridge University Press, 1999.

\bibitem{LasotaMackey}
Andrzej Lasota and Michael~C. Mackey.
\newblock {\em {Chaos, fractals, and noise}}.
\newblock Springer-Verlag, New York, second edition, 1994.

\bibitem{UlamvonNeumann1947}
S.~Ulam and J.~von Neumann.
\newblock On combination of stochastic and deterministic processes.
\newblock {\em Bulletin of the American Mathematical Society}, 53(11):1120,
  1947.

\bibitem{may1976simple}
R.M. May.
\newblock Simple mathematical models with very complicated dynamics.
\newblock {\em Nature}, 261(5560):459--467, 1976.

\bibitem{mcmc_liu}
J.~S. Liu.
\newblock {\em {M}onte {C}arlo Strategies in Scientific Computing}.
\newblock Springer-Verlag, 2001.

\bibitem{KlusKoltaiSchutte2016}
Stefan Klus, P\'{e}ter Koltai, and Christof Sch\"{u}tte.
\newblock On the numerical approximation of the {P}erron--{F}robenius and
  {K}oopman operator.
\newblock {\em Journal of Computational Dynamics}, 3(1):51--79, 2016.

\bibitem{grossmann1977invariant}
Siegfried Grossmann and Stefan Thomae.
\newblock Invariant distributions and stationary correlation functions of
  one-dimensional discrete processes.
\newblock {\em Zeitschrift f{\"u}r Naturforschung a}, 32(12):1353--1363, 1977.

\bibitem{ershov1988solution}
S.~V. Ershov and Georgii~Gennad'evich Malinetskii.
\newblock The solution of the inverse problem for the {P}erron--{F}robenius
  equation.
\newblock {\em USSR Computational Mathematics and Mathematical Physics},
  28(5):136--141, 1988.

\bibitem{diakonos1996construction}
F.K. Diakonos and P.~Schmelcher.
\newblock On the construction of one-dimensional iterative maps from the
  invariant density: The dynamical route to the beta distribution.
\newblock {\em Physics Letters A}, 211(4):199--203, 1996.

\bibitem{diakonos1999stochastic}
FK~Diakonos, D.~Pingel, and P.~Schmelcher.
\newblock A stochastic approach to the construction of one-dimensional chaotic
  maps with prescribed statistical properties.
\newblock {\em Physics Letters A}, 264(2-3):162--170, 1999.

\bibitem{pingel1999theory}
D.~Pingel, P.~Schmelcher, and F.K. Diakonos.
\newblock Theory and examples of the inverse {F}robenius--{P}erron problem for
  complete chaotic maps.
\newblock {\em Chaos}, 9(2):357--366, 1999.

\bibitem{bollt2000controlling}
Erik~M Bollt.
\newblock Controlling chaos and the inverse {F}robenius--{P}erron problem:
  global stabilization of arbitrary invariant measures.
\newblock {\em International Journal of Bifurcation and Chaos},
  10(05):1033--1050, 2000.

\bibitem{nie2013new}
Xiaokai Nie and Daniel Coca.
\newblock A new approach to solving the inverse {F}robenius--{P}erron problem.
\newblock In {\em 2013 European Control Conference (ECC)}, pages 2916--2920.
  IEEE, 2013.

\bibitem{nie2018matrix}
Xiaokai Nie and Daniel Coca.
\newblock A matrix-based approach to solving the inverse {F}robenius--{P}erron
  problem using sequences of density functions of stochastically perturbed
  dynamical systems.
\newblock {\em Communications in Nonlinear Science and Numerical Simulation},
  54:248--266, 2018.

\bibitem{rogers2008synthesis}
A.~Rogers, R.~Shorten, D.M. Heffernan, and D.~Naughton.
\newblock Synthesis of piecewise-linear chaotic maps: Invariant densities,
  autocorrelations, and switching.
\newblock {\em International Journal of Bifurcation and Chaos},
  18(8):2169--2189, 2008.

\bibitem{wei2015solutions}
Nijun Wei.
\newblock Solutions of the inverse {F}robenius--{P}erron problem.
\newblock Master's thesis, Concordia University, 2015.

\bibitem{Ulam1960}
Stanisław~Marcin Ulam.
\newblock {\em A Collection of Mathematical Problems}.
\newblock Interscience Publishers, 1960.

\bibitem{rosenblatt-1952}
M.~Rosenblatt.
\newblock Remarks on a multivariate transformation.
\newblock {\em Ann. Math. Stat.}, 23(3):470--472, 1952.

\bibitem{Gaspard1998}
Pierre Gaspard.
\newblock {\em Chaos, scattering and statistical mechanics}.
\newblock Cambridge University Press, 1998.

\bibitem{Note2}
We have used the language of differential maps, as all the maps that we display
  in this paper are differentiable almost everywhere~\cite {varberg1971change}.
  More generally, $|J(x)|^{-1}$ denotes the density of $\rho _{n} M^{-1}$ with
  respect to $\rho _{n+1}$, see, e.g., \cite [Remark 3.2.4.]{LasotaMackey}.

\bibitem{Note3}
Hence $T^c$ is the translation operator on the unit circle $S^1$.

\bibitem{gyorgyi1984fully}
G~Gy{\"o}rgyi and P~Sz{\'e}pfalusy.
\newblock Fully developed chaotic 1-d maps.
\newblock {\em Zeitschrift f{\"u}r Physik B Condensed Matter}, 55(2):179--186,
  1984.

\bibitem{johnson1987multivariate}
M.E. Johnson.
\newblock {\em Multivariate Statistical Simulation}.
\newblock John Wiley \& Sons, 1987.

\bibitem{devroye-rvgen-1986}
L.~Devroye.
\newblock {\em Non-Uniform Random Variate Generation}.
\newblock Springer-Verlag, 1986.

\bibitem{hormann-rvgen-2004}
W.~H\"{o}rmann, J.~Leydold, and G.~Derflinger.
\newblock {\em Automatic Nonuniform Random Variate Generation}.
\newblock Springer-Verlag, 2004.

\bibitem{dolgov2020approximation}
Sergey Dolgov, Karim Anaya-Izquierdo, Colin Fox, and Robert Scheichl.
\newblock Approximation and sampling of multivariate probability distributions
  in the tensor train decomposition.
\newblock {\em Statistics and Computing}, 30(3):603--625, 2020.

\bibitem{Note4}
When $U$ satisfies the stronger condition that $\protect \operatorname
  {Unif}([0,1]^d)$ is the equilibrium distribution, $U$ is called an exact
  map~\cite {LasotaMackey}.

\bibitem{dick2013high}
Josef Dick, Frances~Y Kuo, and Ian~H Sloan.
\newblock High-dimensional integration: the quasi-{M}onte {C}arlo way.
\newblock {\em Acta Numerica}, 22:133, 2013.

\bibitem{Note5}
The two period sawtooth map $s_2$ is also called the Bernoulli map, and its
  orbit $\protect \mathcal {O}^+(x)$ is the dyadic transformation.

\bibitem{Note6}
This is the `broken linear transformation' in~\cite {grossmann1977invariant} of
  order $p=2\ell $.

\bibitem{rogers2004synthesizing}
Alan Rogers, Robert Shorten, and Daniel~M Heffernan.
\newblock Synthesizing chaotic maps with prescribed invariant densities.
\newblock {\em Physics Letters A}, 330(6):435--441, 2004.

\bibitem{huang2005characterizing}
W.~Huang.
\newblock Characterizing chaotic processes that generate uniform invariant
  density.
\newblock {\em Chaos, Solitons \& Fractals}, 25(2):449--460, 2005.

\bibitem{Note7}
Computation was performed in MatLab that implements IEEE Standard 754 for
  double-precision binary floating-point format.

\bibitem{Note8}
A pre-2006 New Zealand $50$ cent coin.

\bibitem{gentle2003random}
James~E Gentle.
\newblock {\em Random number generation and {M}onte {C}arlo methods}, volume
  381.
\newblock Springer, 2003.

\bibitem{parno2018transport}
Matthew~D Parno and Youssef~M Marzouk.
\newblock Transport map accelerated {M}arkov chain {M}onte {C}arlo.
\newblock {\em SIAM/ASA Journal on Uncertainty Quantification}, 6(2):645--682,
  2018.

\end{thebibliography}

\end{document}